\documentclass[a4paper,11pt]{article}
\usepackage{jheppub} 
\usepackage{tikz}
\usetikzlibrary{arrows.meta, positioning, quotes}
\usepackage{xcolor}
\usepackage{amsmath}
\usepackage{float}
\usepackage{caption}
\usepackage{graphicx}
\usepackage{multirow}
\usepackage{subcaption}
\def\cA{\mathcal{A}}

\title{\boldmath Exploring Accelerating Hairy Black Holes in 2+1 Dimensions: The Asymptotically Locally Anti-de Sitter Class and its Holography}

\author[a,b]{Adolfo Cisterna}
\affiliation[a]{Institute of Theoretical Physics, Faculty of Mathematics and Physics,
Charles University, V Hole{\v s}ovi{\v c}k{\' a}ch 2, 180 00 Prague 8, Czech Republic}
\affiliation[b]{Sede Esmeralda, Universidad de Tarapac\'a, Av. Luis Emilio Recabarren 2477, Iquique, Chile}
\emailAdd{adolfo.cisterna@mff.cuni.cz}

\author[c]{Felipe Diaz}
\emailAdd{f.diazmartinez@uandresbello.edu}
\affiliation[c]{Departamento de Ciencias F\'isicas, Universidad Andres Bello, Sazi\'e 2212, Santiago de Chile}

\author[d,e]{Robert B. Mann}
\emailAdd{rbmann@uwaterloo.ca}
\affiliation[d]{Department of Physics and Astronomy, University of Waterloo, Waterloo, Ontario, Canada, N2L 3G1}
\affiliation[e]{Perimeter Institute, 31 Caroline St. N., Waterloo, Ontario, N2L 2Y5, Canada}

\author[f]{Julio Oliva}
\emailAdd{juoliva@udec.cl}
\affiliation[f]{Departamento de F\'isica, Universidad de Concepci\'on, Casilla, 160-C, Concepci\'on, Chile}

\abstract{
In the realm of lower-dimensional accelerating spacetimes, it is well-established that the presence of domain walls, which are co-dimension one topological defects, is a necessary condition for their construction. We expand upon the geometric framework employed in the generation of such spacetime solutions by incorporating a  conformally coupled scalar field within the matter sector. 
This endeavor leads to the identification of several new families of three-dimensional accelerating spacetimes with asymptotically locally anti-de Sitter (AdS) behavior. Notably, one of these solutions showcases a hairy generalization of the accelerating BTZ black hole. This solution is constructed at both slow and rapid phases of acceleration, and its connection with established vacuum spacetime models is explicitly elucidated. The inclusion of the scalar field imparts a non-constant Ricci curvature to the domain wall, thereby rendering these configurations particularly suitable for the construction of two-dimensional quantum black holes.
To establish a well-posed variational principle in the presence of the domain wall, two essential steps are undertaken. First, we extend the conventional renormalized AdS$_3$ action to accommodate the presence of the scalar field. Second, we explicitly incorporate the Gibbons–Hawking–York term associated with the internal boundaries of our geometries and account for the tension of the domain wall in the action. This dual step process enables us to derive the domain wall field equations via the variational principle. Consequently, the action furnishes the appropriate quantum statistical relation.
We engage in holographic computations, thereby determining the explicit form of the holographic stress tensor. In this context, the stress tensor can be expressed as that of a perfect fluid situated on a curved background. Additionally, it paves the road to ascertain the spacetime mass. Finally, we close by demonstrating the existence of three-dimensional accelerating spacetimes with asymptotically locally flat and asymptotically locally de Sitter geometries, particularly those embodying black holes.}

\begin{document}
\maketitle
\flushbottom

\section{Introduction}

Accelerating black holes are characterized by the line element commonly known, after the preliminary spacetime classification \cite{Ehlers:1962zz}, as the C-metric. Initially discovered by Levi-Civita \cite{levicmetric} and later by Weyl \cite{weylcmetric}, soon after the establishment of General Relativity (GR), the C-metric received no detailed analysis until the work of Kinnersley and Walker \cite{Kinnersley:1969zz} and Bonnor \cite{bonnorcmetric}, who showed that it represents the geometry of two black holes that are causally disconnected and experience acceleration due to the presence of topological defects, specifically cosmic strings that pull (or struts that push) the black holes away from each other. This extension of the Schwarzschild spacetime, governed by one extra parameter, has been extensively studied in the context of GR \cite{Letelier:1998rx, Bicak:1999sa, Podolsky:2000at, Pravda:2000zm, Dias:2002mi, Griffiths:2005qp, Krtous:2005ej} as well as in other scenarios, including Einstein-dilaton-Maxwell \cite{Dowker:1993bt}, braneworld models \cite{Emparan:1999wa},  quantum black holes \cite{Emparan:1999fd, Emparan:2000fn, Gregory:2008br, Emparan:2020znc}, and (gauged) supergravities \cite{Lu:2014ida, Lu:2014sza, Nozawa:2022upa}. Recently, there have been proposals to test the astrophysical implications of such objects
\cite{Grenzebach:2015oea,Ashoorioon:2022zgu,Ashoorioon:2021gjs,Zhang:2020xub}.

The C-metric spacetime possesses axial symmetry, and a detailed analysis of the axis of symmetry reveals the existence of conical singularities. These conicities appear at both the north and south poles of the symmetry axis and are characterized by either an angular deficit or an angular excess. However, by appropriately defining the period of the azimuthal angle, it is possible to mitigate the conicity at one of the poles, typically the north pole, with the entire conicity manifest as an angular deficit at the south pole. In physical terms, this conicity is attributed to a topological defect, a cosmic string extending from $r = 0$ to conformal infinity. The string exerts a force on the black holes, causing their acceleration.
Due to the specific geometric properties of the C-metric line element, this spacetime is primarily limited to four dimensions, with a notable exception in the three-dimensional case.
In this lower-dimensional scenario, two significant differences arise: First, without an additional angular dimension, there is no room to accommodate conical singularities, necessitating a reinterpretation of the mechanism responsible for the black holes' acceleration. Second, the angular coordinate, on which the metric functions depend, ceases to be a polar coordinate (as in the four-dimensional case) and becomes an azimuthal angle whose identification becomes crucial for determining the causal structure of the solution. It is this identification of the three-dimensional angular coordinate that leads to the existence of a compact event horizon.

Three-dimensional gravity is known to be topological and it can be considered trivial at the classical level, i.e. there are no gravitational waves. In addition, the theory is non-renormalizable by power counting, implying its incapability to be quantized. 
Notwithstanding this, it has been shown \cite{Achucarro:1986uwr, Witten:1988hc} that three-dimensional gravity can be made equivalent to a Chern--Simons gauge theory if the vierbein and the spin connection are properly combined into a gauge valued connection where the gauge group depends on the value of the cosmological constant. 
Therefore, pure gravity in three dimensions is a gauge theory; this is in contrast with the four-dimensional case, where if one writes an $ISO(3,1)$ valued connection in terms of the vierbein and the spin connection, the resulting action is not a gauge theory. 
It was shown that the theory has a well-defined renormalizable perturbation expansion with vanishing beta function \cite{Witten:1988hc}. Moreover, in \cite{Witten:1988hf} Witten has shown that the partition function of Chern--Simons theory is equivalent to the Jones polynomial, where a Wilson loop specified by the given knot associated with the polynomial is found\footnote{This extends also to four-dimensional super Yang--Mills theory where the observables are related to the Khovanov homology of knots \cite{Witten:2011zz}.}. For a negative cosmological constant, Witten also provided a relation between the quantization of three-dimensional gravity and a certain two-dimensional conformal field theory.
These results, together with the seminal work of Brown and Henneaux \cite{Brown:1986nw}, who showed that the constraint algebra of AdS$_3$ gravity, upon taking a suitable set of boundary conditions, gives rise to a Virasoro algebra with computable central charge, in some sense an early version of the subsequently introduced AdS/CFT correspondence conjecture. 

Three-dimensional gravity is therefore simple, but yet rich enough, to study a quantum theory of gravity: in particular, the Hilbert space of gravity with a negative cosmological constant, which corresponds to the space of conformal blocks in two dimensions. This matches with a three-dimensional version of the celebrated AdS/CFT correspondence \cite{Maldacena:1997re, Gubser:1998bc, Witten:1998qj}. 
 Due to the topological nature of the theory, all solutions are locally equivalent to the AdS.  Despite this, a black hole solution was obtained by 
 Ba\~nados, Teitelboim, and Zanelli
 and is known as the BTZ black hole \cite{Banados:1992wn}. This black hole features an $S^1\times T^2$ topology in the Euclidean continuation and corresponds \cite{Banados:1992gq} to a coset space of AdS$_3$ and a discrete subgroup, acting discretely on the automorphism group $SO(3,1)$. The latter is a feature of all solutions of three-dimensional AdS gravity.
 
Using modular bootstrap techniques, the sum over all known contributions to the Euclidean partition function was computed
\cite{Maloney:2007ud, Keller:2014xba,Benjamin:2019stq}, 
showing that the resulting partition function contains an infinite tower of negative states corresponding to a non-unitary partition function with no modular invariance. This is inconsistent within the realm of the AdS/CFT correspondence and indicates that the quantum theory may not exist. 
Some attempts to solve this puzzle involve including topologies that do not have a semiclassical limit \cite{Maxfield:2020ale}, receding from pure gravity to include non-smooth boundaries \cite{Benjamin:2020mfz}, or considering states with spin scaling corresponding to strongly coupled strings in the bulk \cite{DiUbaldo:2023hkc}. 
 A very appealing proposal to cure this problem, proposed by Maloney and Witten, entails the inclusion of matter fields, scalars or other fields arising from dimensional compactification of superstring models \cite{Maloney:2007ud}. A particular example of this considers type IIB superstring theory with quantized fluxes\footnote{The fluxes take quantized values of the ratio $\ell/G$~, which correspond to the moduli parameters of AdS gravity. In four dimensions this implies a problem in perturbation theory \cite{Witten:2007kt}, but for the three-dimensional case this issue is mapped to the fact that the holographic central charge depends on this ratio;  according to the c-theorem \cite{Zamolodchikov:1986gt} this must be a constant. This indicates that the quantized values of the fluxes correspond to the critical points in which the dual CFT admits a holomorphic factorization.} determined by the compactification to AdS$_3$~, with domain walls across which the fluxes jump. 
 An interesting spacetime of pure AdS$_3$ gravity satisfying these conditions is the one representing accelerating three-dimensional black holes \cite{Astorino:2011mw, Arenas-Henriquez:2022www}, with the acceleration sourced by a domain wall topological defect.

Recent research \cite{Arenas-Henriquez:2022www} has demonstrated how to construct accelerating black hole solutions with compact horizons by employing appropriate identifications of the angular coordinate\footnote{An accelerating generalization of three-dimensional static black holes was early provided in \cite{Astorino:2011mw} and later extended in \cite{Xu:2011vp}.}. These identifications are made possible by introducing a topological defect of co-dimension one, referred to as a domain wall, which plays a crucial role in accelerating these three-dimensional black holes. The domain wall replaces the conical singularities encountered in four dimensions and is an essential element in establishing a well-posed action principle and a regularizable action \cite{Arenas-Henriquez:2023hur}. The domain wall has a contribution to the partition function already at the saddle point.

The inherent complexities of the C-metric geometry have limited the number of known solutions in the presence of matter, with most of these solutions relying on ad hoc symmetries of the matter fields and being confined to four dimensions \cite{Charmousis:2009cm,Anabalon:2009qt,Barrientos:2022bzm,Hale:2023qjx,Nozawa:2023pzn}. The exploration of the three-dimensional  C-metric geometry remains largely uncharted, with the exception of a solution incorporating power-Maxwell electrodynamics \cite{EslamPanah:2022ihg}. Here we endeavor to address this gap by presenting, for the first time, three-dimensional black hole solutions featuring conformal scalar hair within the framework of the C-metric geometry. In doing so we give an explicit setting in which the proposal of Maloney and Witten can be concretized. 
We provide a self-contained guide to go through an extensive family of black hole solutions with conformal scalar hair that generalize the vacuum solutions presented in \cite{Astorino:2011mw,Xu:2011vp,Arenas-Henriquez:2022www}.

We initiate our investigation by demonstrating the integration of the solutions and establishing in appropriately chosen coordinates, the location of the Killing horizons. Furthermore, we explore their relative positioning in relation to conformal infinity, while also portraying the causal structure inherent in the solutions. In order to achieve this objective, we adhere to the approach outlined in \cite{Arenas-Henriquez:2022www}, wherein we define the range of coordinates and glue two copies of the spacetime by incorporating a domain wall. In addition, restrictions imposed by requiring an everywhere real and well-behaved scalar field profile are also taken into account. These constructive steps enable us to obtain compact event horizons. Additionally, we thoroughly examine the global aspects of these geometries, while also conducting holographic computations to ascertain the spacetime mass.\\

This work is organized as follows: Section II provides preliminary results, specifically, a pedagogical introduction to three-dimensional accelerating black holes in vacuum. In Section III we start by presenting our model, the field equations, and the integration of the solutions. Next, we introduce our scheme indicating how the coordinates are to be restricted and identified, via the inclusion of a domain wall, in order to construct physically meaningful spacetimes. The corresponding junction conditions are explicitly shown for our scalar-tensor model. We close this section by providing two families of asymptotically locally AdS (AlAdS) three-dimensional hairy accelerating black holes. In Section IV we conduct holographic computations. We show how to build a well-defined action principle and how to compute the corresponding holographic stress tensor. Then, we compute the black hole mass. Section V is devoted to presenting interesting solutions with asymptotically locally de Sitter (dS) and asymptotically locally flat behavior, while Section VI is destined to provide our final conclusions with a particular emphasis on future research directions that emerge from this work. Appendix A delivers a qualitative analysis of the Killing horizons of our solutions.  

\section{Preliminaries}

To comprehensively examine the solutions presented in this work, it is instructive to first delve into the construction of solutions for the vacuum case and briefly review the spectrum of solutions obtained therein. After performing the integration of the field equations, two crucial steps need to be taken. First, it is essential to precisely define the range of spacetime coordinates, particularly the transverse coordinate, which is vital for preserving the correct maintenance of the spacetime metric signature and determining the number of Killing horizons present in the geometry. Second, after establishing the range of the aforementioned coordinate, a mirroring of the spacetime becomes necessary. This mirrored copy is then carefully attached to the original geometry, resulting in solutions characterized by compact event horizons. This gluing process is effectively achieved by introducing a topological defect, specifically a domain wall of codimension one. The approach adopted in this section follows the analysis conducted in \cite{Arenas-Henriquez:2022www}. Throughout this section, we will exclusively employ prolate coordinates as they significantly simplify the analysis of the spacetime's causal structure.

\subsection{2+1 accelerating black holes in Einstein gravity}\label{Sec:vacuumsolutions}

We commence our discussion by introducing the utilization of prolate coordinates $(t,y,x)$~, with the corresponding ranges appropriately defined for specific spacetime scenarios. As a result, the $(2+1)$-dimensional line element can be expressed as follows
\begin{equation}
ds^2=\frac{1}{A^2(x+y)^2}\left[-F(y)dt^2+\frac{dy^2}{F(y)}+\frac{dx^2}{G(x)}\right]~. \label{prolateansatz}
\end{equation}
The parameter $A$ is known to be associated with the acceleration of a Rindler observer. Notice that the region $x + y = 0$ defines a conformal boundary. The fact that conformal infinity is given by a non-constant surface rather than a point (usually at radial infinity in polar coordinates) is a prominent feature of accelerating solutions. 

The metric polynomials can be readily deduced from the trace of the field equations \cite{Plebanski:1976gy}, wherein the existence of a non-vanishing cosmological constant $\ell^{-2}:=-\Lambda$ yields the following expressions
\begin{align}\label{polynomialsvacuum}
F(y)&=y_3y^3+y_2y^2+y_1y+y_0+\frac{1}{A^2\ell^2}~,\nonumber\\ 
G(x)&=x_3x^3+x_2x^2+x_1x+x_0~,
\end{align}
where the condition $x_3=y_3$ is  necessary. Subsequently, upon substituting these polynomials into the remaining field equations,  yield the additional constraints $y_2 = -x_2$~, $y_1 = x_1$~, and $y_0 = -(\Lambda+A^2x_0)/A^2$~,
and the cubic contributions, namely $x_3=y_3$~, vanish.
Consequently, the solution reduces to a pair of quadratic polynomials in the $y$ and $x$ coordinates\footnote{As three-dimensional gravity is topological, the metric \eqref{prolateansatz} can be cast in pure AdS by applying coordinates transformations, see \cite{Arenas-Henriquez:2022www}.}. 

By exploiting the evident symmetry of the line element $(t\rightarrow st~, y\rightarrow y-s~, x\rightarrow x+s)$ with $s$ as a constant, combined with the reparametrization $A\rightarrow{sA}$~, we are able to entirely eliminate redundancies stemming from the arbitrariness of $x_{2}$ and $x_{1}$~. As a result, only sign differences\footnote{When $x_2 = 0$, the metric function $F(y)$ becomes a constant, resulting in the absence of Killing horizons. This particular solution corresponds to the massless BTZ spacetime  in the zero-acceleration limit, as discussed in more detail in \autoref{SubSec:AdSsols}.} among the preceding polynomial coefficients retain significance, leading to the identification of three distinct forms of the spacetime. These forms are succinctly summarised in \autoref{table1}.

\begin{table}[H]
\centering
\begin{tabular}{ | c || c | c | c | }
    \hline
    Class   & $G(x)$    & $F(y)$                  & Maximal range of $x$
    \\
    \hline\hline 
    I       & $1-x^2$   & $\frac{1}{A^2\ell^2}+(y^2-1)$ & $|x|<1$
    \\
    II      & $x^2-1$   & $\frac{1}{A^2\ell^2}+(1-y^2)$ & $x>1$ or $x<-1$
    \\
    III     & $1+x^2$   & $\frac{1}{A^2\ell^2}-(1+y^2)$ & $\mathbb{R}$
    \\
    \hline 
\end{tabular}
\caption[
    Metric functions for the three-dimensional C-metric
  ]{Three distinct classes of solutions, each exhibiting a specific maximal range for the transverse coordinate, are identified. It is important to note that a potential linear term has been effectively eliminated through a coordinate transformation. The coordinate pair $(t,y)$ spans the entire real line, ranging from $(-\infty,\infty)$~, while the range of $x$ is constrained to ensure the preservation of the correct metric signature.} 
\label{table1}
\end{table}
At this point, the employment of prolate coordinates proves to be advantageous, as it offers a swift means to impose constraints on the coordinates, ensuring the preservation of the metric signature. Solutions belonging to Class II were initially discovered in \cite{Astorino:2011mw}, and more recently, an in-depth investigation of these geometries was carried out in \cite{Arenas-Henriquez:2022www}. \autoref{table1} provides an overview of three families of solutions: Class I primarily represents spacetimes of accelerating particle-like solutions, characterized by naked conical singularities encompassed solely by a Rindler horizon. A specific region within its parameter space reveals the existence of an accelerating black hole solution that lacks a continuous limit to the standard BTZ geometry, a configuration referred to as Class I$_{\rm C}$~.
In contrast, Class II constitutes a one-parameter extension of the BTZ static black hole, possessing a well-defined vanishing acceleration limit, a feature absent in the BTZ extension of Class I$_{\rm C}$~. Finally, Class III does not encompass particle-like solutions or accelerating black holes by merely performing the gluing along one domain wall. Instead, it involves a more intricate topological structure, which is beyond the scope of interest in the present paper.

\begin{figure}[h]
    \centering
    \includegraphics[scale=1]{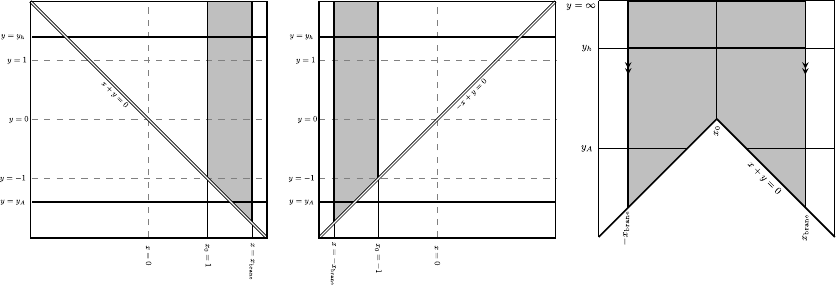}
    \caption{The causal structure of the rapidly accelerating BTZ black hole can be elucidated as follows. In the initial depiction (left figure), we establish the range of coordinates and ascertain the positions of the Killing horizons. Moving forward, the middle figure illustrates the mirroring of the first one, and the last figure showcases the ultimate causal structure of the spacetime, achieved by gluing together, at $x_0$ and $x_{\rm brane}$~, the configurations from the left and middle figure. $x_0$ remains as a tensionless locus, while $x_{\rm brane}$ represent the position of the domain wall.}
    \label{fig:side_by_side1}
\end{figure}

In order to provide an illustrative example, we shall briefly examine the geometry belonging to Class II, focusing specifically on the scenario where $x>1$~. To proceed, we define an appropriate range for the $x-$coordinate as $(x_{\rm min}, x_{\rm max})$~, with the constraint that $x_{\rm min}>1$~. Subsequently, we investigate the location of the Killing horizons, which correspond to the zeros of the metric function $F(y)$. These horizons are located at
\begin{equation}
y_A=\frac{\sqrt{1+A^2\ell^2}}{A\ell}~,\hspace{0.3cm}y_h=-\frac{\sqrt{1+A^2\ell^2}}{A\ell}~.
\end{equation}
Here, $y_A$ and $y_h$ represent the accelerating and event horizons, respectively. We refine the constraint on $x_{\rm max}$ based on two distinct conditions: $x_{\rm max} > y_A$ for the rapid acceleration phase, or $x_{\rm max} < y_A$ for the slow acceleration phase. These conditions delineate whether the black hole exhibits a single compact Killing horizon or also a second, non-compact Rindler horizon. This information enables us to delineate the causal structure of the solution, depicted in \autoref{fig:side_by_side1}. 
To construct the compact event horizon, we mirror the spacetime depicted in the left of \autoref{fig:side_by_side1} along two points dubbed $x_{0}$ and $x_{\rm brane}$~, which must belong to the set $(x_{\rm min}, x_{\rm max})$~. This yields the middle figure of \autoref{fig:side_by_side1}. 
Subsequently, we unite the mirrored copy with the original geometry by introducing a domain wall. The final composite spacetime is shown on the right. The domain wall possesses a negative tension, which, through Israel junction conditions, is determined to be $\mu=-2A\sqrt{G(x_{\rm brane})}/\kappa$~.
Observe that due to the form of the tension, only one domain wall is required. This comes from the fact that we can first glue the two copies at $x_{0}$ representing a root of the metric polynomial $G(x)$~, therefore providing a tensionless locus. The second glue point is given by $x_{\rm brane}$~, the actual localization of a tension-full domain wall.  
It is important to note that we have constructed a scenario where both Rindler and event horizons are present, and specifically, for the case where $x>1$~. Consequently, this configuration represents a rapidly accelerating extension of the BTZ black hole, accelerated by a strut characterized by a negative domain wall tension. 

With the framework established, the subsequent sections of this paper will focus on constructing and analyzing scalar hairy extensions of the vacuum geometries previously presented in \cite{Astorino:2011mw,Arenas-Henriquez:2022www}. These extensions will involve the inclusion of a conformally coupled scalar field in the spacetime, and through rigorous analysis, we aim to investigate their influence on the overall geometry and physical properties of the resulting solutions.

\section{2+1 accelerating black holes with conformal hair}

\subsection{Model, field equations, and integration of the solutions}

We enhance Einstein-$\Lambda$ theory by incorporating the action of a self-interacting conformally coupled scalar field.
Conformally coupled scalar fields have found applications in modeling quantum effects within semiclassical theories of gravity \cite{PhysRevD.21.2756, PhysRevD.32.1302, Phillips:1996em}. 
    Notably, in two dimensions \cite{Kolanowski:2023hvh} and three dimensions \cite{Hubeny:2009rc, Emparan:2020znc}, holography has proven instrumental in addressing the challenge of quantum backreactions of conformal scalars. This approach offers a non-perturbative solution to the study of quantum effects in the context of black holes. Furthermore, it's worth noting that the theory featuring conformally coupled scalar fields exhibits a well-posed Cauchy problem \cite{Noakes:1983xd}. This choice also renders the description of wave propagation in curved backgrounds more well-defined when compared to using minimally coupled scalars \cite{Sonego:1993fw}.

    In the context of classical black hole physics, the well-known no-hair theorems \cite{Ruffini:1971bza, Bekenstein} have been circumvented by considering conformally coupled scalar fields \cite{ bocharova1970exact, BEKENSTEIN1974535}.
    In three dimensions a black hole dressed with a conformal scalar was found in \cite{Martinez:1996gn}, and subsequently generalized in \cite{Henneaux:2002wm} by considering a hexic self-interacting potential, which preserves the conformal symmetry of the matter sector. These solutions exhibit asymptotically locally Anti-de Sitter (AdS) behavior and have been extensively studied in the literature (see for instance \cite{Martinez:2009kua}, and reference therein). Additionally, it is worth noting that the scalar field profiles in these solutions exhibit a weak asymptotic decay, which results in non-standard contributions to the Hamiltonian charges and boundary variations \cite{Gegenberg:2003jr}. Therefore, the conserved quantities associated with these solutions deviate from those of the conventional BTZ black hole. The canonical generators of asymptotic symmetries are also modified by the presence of the conformal scalar but relaxing consistently \cite{Henneaux:2002wm} the Brown--Henneaux boundary conditions \cite{Brown:1986nw} such that the asymptotic symmetry algebra still corresponds to the Virasoro algebra with the same central charge. 

The corresponding action in three dimensions we shall work with reads
\begin{equation}
I[g_{\mu\nu},\phi]=\int_{\cal M} d^3x \sqrt{-g}\left[\frac{R-2\Lambda}{2\kappa}-\frac{1}{2}\left(\partial\phi\right)^2-\frac{1}{16}R\phi^2-\lambda\phi^6\right]~. \label{action1}
\end{equation}
 Newton's constant is denoted by $\kappa=8\pi G$~, and the dimensionless parameter $\lambda$ governs the conformal potential, which is hexic in three dimensions. 
 By performing stationary variations with respect to the fields, we obtain the field equations
\begin{align}
E_{\mu\nu} :={}& G_{\mu\nu}+\Lambda g_{\mu\nu}-\kappa T_{\mu\nu} =0~,\\
0 ={}& \square\phi-\frac{1}{8}R\phi-6\lambda\phi^5~, \label{EOM}
\end{align}
where we have defined the improved energy-momentum tensor
\begin{equation}
     T_{\mu\nu} = \partial_\mu\phi\partial_\nu\phi-\frac{1}{2}g_{\mu\nu}\left(\partial\phi\right)^2+\frac{1}{8}\left(G_{\mu\nu}-\nabla_\mu\nabla_\nu+g_{\mu\nu}\square\right)\phi^2-g_{\mu\nu}\lambda\phi^6~.
\end{equation}
The integration of the system is carried out smoothly by employing prolate coordinates, wherein the metric ansatz \eqref{prolateansatz} is complemented by a scalar field configuration of the form $\phi(y,x)$~. The specific range of coordinates will be determined once particular geometries are constructed, following a similar approach as in the vacuum case. The trace of the field equations remains free from any scalar field contribution, which arises as a consequence of the Weyl-rescaling invariance of the scalar field action. As a result, the metric polynomials maintain the same form as previously presented in \eqref{polynomialsvacuum}.
Meanwhile, the scalar field profile can be obtained from a suitable combination of equations, specifically $E^{t}_t-E^y_y=0$~, resulting in the expression
\begin{equation}
\phi(y,x)=\sqrt{-\frac{x+y}{F_1+(x+y)F_2}}~. \label{scalar}
\end{equation}
The functions $F_1=F_1(x)$ and $F_2=F_2(x)$ are two arbitrary functions of the $x$-coordinate, whose forms will be determined by the remaining field equations. The final solution is obtained by inserting the polynomials given in \eqref{polynomialsvacuum}, combined with \eqref{scalar}, into the field equations. This procedure ultimately leads to the determination of the functions $F_1$~, $F_2$~, and the polynomial coefficients $x_i$ and $y_i$ that characterize the spacetime geometry and the scalar field configuration.
It is important to note that, in contrast to the vacuum case, the field equations will now involve non-vanishing cubic coefficients, which arise as a consequence of the Weyl invariance of the matter sector being considered. 
This distinction marks a significant departure from the vacuum scenario and highlights the interplay between the gravitational and scalar field components in the context of the extended theory. 

While it is straightforward to obtain a hairy extension of the vacuum solutions  \cite{Astorino:2011mw,Xu:2011vp,Arenas-Henriquez:2022www} using prolate coordinates, 
it is more advantageous to employ a geometric gauge that aligns with our experience in dealing with conformally coupled matter sources 
to study the causal structure of the spacetime.
In Maxwell's theory, it is well-known that the backreaction induced by monopole electric and magnetic charges complements the non-Newtonian term that arises from solving the trace of the field equations. To illustrate this,  consider the higher dimensional conformal extension of Maxwell theory and, for specificity, a static spherically symmetric configuration in which the line element exhibits a single metric function $f(r)$~. The trace of the field equations then implies a backreaction, yielding the form
\begin{equation}
    f(r)=1-\frac{A}{r^{d-3}}+\frac{B}{r^{d-2}}+\frac{r^2}{\ell^2}~.
\end{equation}
This backreaction precisely coincides with the one obtained by solving all the field equations of the Einstein-Maxwell (conformal) system, with the appropriate identification of the integration constants $A$ and $B$ to physically meaningful parameters \cite{hassainemartinez}. A similar observation holds true for other Weyl-rescaling invariant matter sources, such as conformally coupled scalar fields. Consequently, in our $2+1$-accelerating case, we expect our line element to acquire a cubic contribution in the metric polynomials, analogous to a $\sim 1/r$ term in spherical coordinates. Consequently, our hairy configuration should bear a striking resemblance to the standard four-dimensional C-metric spacetime\footnote{The understanding of this can be explained as follows. In the spherically symmetric case, the integrated metric function $f$ derived from the trace equation exhibits the form $f\sim r^2/\ell^2-B/r+A$~. The presence of the $B$ term is consistent with a Weyl-rescaling invariant matter sector, which implies that the entire solution adopts the causal structure of a four-dimensional Schwarzschild (A)dS black hole.}.
Indeed, the field equations \eqref{EOM} admit the following solution
\begin{align}
    F(y) &=-\sigma(1-y^2)(1-\xi y)+\frac{1}{A^2\ell^2}~, \nonumber\\
    G(x) &= \sigma(1-x^2)(1+\xi x)~, \label{SOL}\\
    \phi(y,x)&=\sqrt{\frac{8}{\kappa}}\sqrt{\frac{x+y}{y+\alpha}}~, \nonumber
\end{align}
where the parameters $\xi$ and $\alpha$ are related via
\begin{equation}
    2\alpha - \xi\left(1-3\alpha^2\right)=0~, \label{constraint1}
\end{equation}
and where consistency with all field equations requires 
\begin{equation}
    \left(1-3 \alpha ^2\right) \left(\Lambda\kappa ^2 +512 \lambda\right)+ \sigma A^2 \kappa ^2\left(1-\alpha ^2\right)^2 = 0~. \label{constraint2}
\end{equation}
$\sigma=\pm1$ allows us to recover the classes I and II of the bald cases ($\xi=0$) presented in \autoref{table1}. Note that the arbitrary functions $F_1$ and $F_2$~, which define the scalar field profile, have already been fixed by the field equations.
In the four-dimensional case, this line element is known as the Hong-Teo gauge for the C-metric spacetime \cite{Hong:2003gx}, and it significantly simplifies the study of the causal structure of accelerating metrics.

A remarkable difference with the vacuum case is the presence of a curvature singularity. As a matter of fact, our solutions exhibit a curvature singularity as $y$ approaches infinity, indicating that they do not represent constant curvature spacetimes, as evident from the Kretschmann invariant
\begin{equation}
\mathsf{K}:=R^{\mu\nu\rho\sigma}R_{\mu\nu\rho\sigma}=6A^4\xi^2(x+y)^6+\frac{12}{\ell^4}~.
\end{equation} 
We will subsequently introduce spherical-like coordinates (refer to \eqref{rescaling}), and this observation corresponds to the emergence of a curvature singularity at $r=0$~, which must be dressed by the Killing horizons of the corresponding geometries. 

The general family of solutions presented here, for the first time, represents an exact hairy generalization of the accelerating metrics found in \cite{Astorino:2011mw,Xu:2011vp,Arenas-Henriquez:2022www}. Some of these solutions correspond to a hairy extension of the accelerating BTZ black hole. Furthermore, due to the presence of the acceleration parameter, several families of stealth black holes can also be constructed. These stealth black holes can serve as seeds that, through appropriate conformal transformations, amplify the spectrum of new hairy accelerating backreactions.

\subsection{Coordinate identifications and junction conditions}\label{SubSec:horizons}

We now proceed with the analysis of the solutions contained in \eqref{SOL}, specifically those forecasting an anti-de Sitter asymptotic behavior. We shall focus on the physically sensible spacetimes that satisfy the condition $x+y>0$. The Lorentzian signature of the solutions is guaranteed by requiring $G(x)>0$. Recall that, in this hairy scenario, the backreaction of the scalar field modifies not only the metric function $F(y)$, but also the function $G(x)$ through the inclusion of the parameter $\xi$. Then, for each value of $\sigma$, there are two possibilities for ensuring $G(x)>0$, namely that each of its factors are simultaneously positive or simultaneously negative, depending on the value of the scalar field constant $\xi$. The possible ranges of the $x-$coordinate for each solution are succinctly summarised in \autoref{table2}, where have focused on the case in which the parameter $\xi$ is positive. This is a simplicity assumption, that indeed can be deleted in order to enlarge the spectrum of solutions contained in \eqref{SOL}.

\begin{table}[H]
\centering
  \begin{tabular}{|c||c||c|}
    \hline
    Class & $\xi>1$ & $0<\xi < 1$ \\
    \hline\hline
    \multirow{2}{*}{$\sigma = 1$} & $x<-1$ & $x<-1/\xi$ \\
    \cline{2-3}
    & $-1/\xi < x < 1$ & $-1<x<1 $\\
    \hline\hline
    \multirow{2}{*}{$\sigma=-1$} & $-1<x<-1/\xi$ & $\emptyset$\\
    \cline{2-3}
    & $x>1$ & $x>1$\hspace{0.1cm}or\hspace{0.1cm}$-1/\xi < x < -1$ \\ \hline
  \end{tabular}
\caption[
    Metric functions for the three-dimensional hairy C-metric
  ]{Domains of the transverse coordinate depending on the possible values of $\xi$ and $\sigma$~. On each family, the first row expresses the result when both factors of $G(x)$ are asked to be negative. On the contrary, the second row expresses the cases in which both factors of the polynomial are positive.} 
\label{table2}
\end{table}

We now analyze the behavior of the scalar field. It can be seen from \eqref{SOL} that the scalar field possesses a pole at $y=-\alpha$~. Consequently, its profile becomes imaginary unless we impose the restriction $y>-\alpha$~. 
From \eqref{constraint1} we obtain two branches  for $\alpha$  
\begin{equation}
\alpha_{\pm}=\frac{-1\pm\sqrt{1+3\xi^2}}{3\xi}~,
\end{equation}
and so the interval of possible values of $\alpha_{\pm}$ depends on the value of $\xi$; these  are
displayed in \autoref{table3}.

\begin{table}[H]
\centering
  \begin{tabular}{|c||c||c|}
    \hline
     & $\xi>1$ & $0<\xi < 1$ \\
    \hline\hline
     $\alpha_+$ & $\alpha_+\in\left(\frac{1}{3},\frac{\sqrt{3}}{3}\right)$ & $\alpha_+\in\left(0,\frac{1}{3}\right)$ \\
    \hline\hline
    $\alpha_-$ & $\alpha_-\in\left(-1,-\frac{\sqrt{3}}{3}\right)$ & $\alpha_-\in\left(-\infty,-1\right)$\\
 \hline
  \end{tabular}
\caption{Possible values of $\alpha$~.} 
\label{table3}
\end{table}
The net effect of this restriction is that of ulterior constraints that must be applied to the $y$ coordinate. Once the value of $\alpha$ is given, the $y$ coordinate must be such that $y>-\alpha$~. We therefore expect to localize the corresponding Killing horizon following this restriction.   
\\

As explained in \cite{Arenas-Henriquez:2022www}, (see \autoref{Sec:vacuumsolutions}) accelerating black hole solutions in three-dimensional gravity are constructed by means of the following steps:
\begin{itemize}
    \item[i)] Once a given solution \eqref{prolateansatz} is found, it is necessary to establish the range of the transverse coordinate $x$ that maintains the correct Lorentzian signature of the metric tensor, i.e $G(x)>0$~. See \autoref{table2}.

    \item[ii)] Next, the values of $x_{\rm min}$ and $x_{\rm max}$ of the transverse coordinate are decided on the basis of how many Killing horizons the desired geometry is allowed to contain. This is complemented by requiring the condition $y>-\alpha$~, if it applies.
 
    \item[iii)] Having at hand the interval defined by $x_{\rm min}$ and $x_{\rm max}$~, and the possible restriction of the $y$ coordinate induced by the reality condition of the scalar field, we produce a mirror copy of the spacetime. We glue it to the original by identifying both copies at $x_0$ and $x_{\rm brane}$~, both locus belonging to the interval $(x_{\rm min},x_{\rm max})$~.
    The gluing is made possible by the introduction of a domain wall topological defect of co-dimension one. 

\end{itemize}

The scheme provides the necessary conditions to construct spacetimes with compact event horizons. Two domain walls $\Sigma_i$ are introduced at $x_i = \{x_0,x_{\rm brane}\}$~, of which the corresponding energy densities $\mu_i$ are given by the Israel junction conditions. Indeed, there is a jump of the extrinsic curvature along these surfaces. The energy density of the walls is proportional to the function $\sqrt{G(x)}$ evaluated at $x_0$ and $x_{\rm brane}$~, respectively. 
In the vacuum case  \cite{Arenas-Henriquez:2022www}   the lower bound  $x_0 = 1$ was chosen, resulting in a vanishing tension for the corresponding domain $\Sigma_{x_0}$; the acceleration is produced by a domain wall located at $x_{\rm brane}$ only. Recall that not all solutions describe black holes. The process can potentially produce accelerating particle-like solutions with a Rindler (non-compact) horizon, depending on the possible values of $x_{\rm brane}$ and the original class (value of $\sigma$) chosen.

We proceed now with the analysis of our hairy configurations. In order to produce configurations with compact event horizons we need to introduce, as mentioned before,  two domain walls at $x_i = \{x_0,x_{\rm brane}\}$~, of which the corresponding line elements are given by 
\begin{align}
    ds^2 = \gamma_{MN}dx^M dx^N = \frac{1}{A^2(x_i +y)^2}\left(-F(y)dt^2 + \frac{dy^2}{F(y)}\right)~.
\end{align}
Here, $(M, N) = (t, y)$ represents the domain walls coordinates.  The scalar curvature of these domain walls
\begin{align}
    {\cal R}^{MN}_{~~~~MN} = 2 A^2 \sigma  \left(\xi  \left(3 x_i^2 y+3 x_i y^2+x_i+y^3\right)-x_i^2+1\right)+2\Lambda~,
\end{align}
is not constant, and
exhibits a curvature singularity at $y\to \infty$~.  This is an appealing feature that opens the possibility for studying two-dimensional quantum black holes\footnote {See for instance \cite{Witten:1991yr} for an analysis of black holes in string theory.} localized on the brane, as it has been done in \cite{Anber:2008zz, Emparan:2020znc,Emparan:2021hyr, Emparan:2022ijy}, where the four-dimensional C-metric is used to study the quantum version of the BTZ black hole. Notice that this is an exclusive feature of our hairy configurations --   the limit $\xi\rightarrow0$ yields a constant scalar curvature.

The extrinsic curvature associated with each domain wall is defined as  ${\cal K}_{MN}:=\frac12 {\cal L}_n \gamma_{MN}$~, where $n^\mu$ is the outward pointing normal of the surface located at $x = x_i$ 
\begin{align}
    n^\mu = \frac{G(x)^{-\frac12}}{A(x+y)}\left(\frac{\partial}{\partial x}\right)^\mu\Bigg\rvert_{x_i}~,
\end{align}
which yields
\begin{align}
    {\cal K}_{MN} = AG(x_i)^{\frac12}\gamma_{MN}~.
\end{align}
The energy density of each wall is obtained along the lines of \cite{Aviles:2019xae}, where the junction conditions for scalar-tensor theories of the type considered here have been described. 
The standard Israel junction conditions get modified by the presence of the conformally coupled scalar field, and the domain wall stress tensor now reads
\begin{align}\label{JC}
T_{MN} = -2\epsilon f(\phi)\left([{\cal K}_{MN}] - \gamma_{MN}[{\cal K}]\right) + 2\epsilon n^\mu [\partial_\mu \phi]  f'(\phi)\gamma_{MN}~,  
\end{align}
which is subject to 
\begin{align}\label{JC2}
    \epsilon n^\mu [\partial_\mu \phi] = 2\epsilon f'(\phi)[{\cal K}]~,
\end{align}
where $[X]:=X^+ - X^-$ corresponds to the difference of a quantity $X$ evaluated on each side of $\Sigma_i$~, $\epsilon = \pm1$ corresponds to the cases where the wall is a timelike or 
spacelike hypersurface respectively and $f(\phi)$ represents the coupling between the scalar field and the Ricci scalar in the action evaluated along $\Sigma_i$~. In our case the latter is given by $f(\phi) = \frac{1}{2\kappa}\left(1-\frac{\kappa}{8}\phi^2\right)$ as can be seen from \eqref{action1}.
Combining \eqref{JC} and \eqref{JC2} we find the wall energy density to be 
\begin{align}
\frac{\kappa}{2}\gamma_{MN} \mu= \frac{\kappa}{2}\int_-^+ T_{MN} = A\sqrt{G(x_i)}\gamma_{MN}~,
\end{align}
which develops the same structure as the energy density of the domain wall of the vacuum accelerating solutions \cite{Arenas-Henriquez:2022www}.  However, in this case the function $G(x) = \sigma(1-x^2)(1+\xi x)$ has an extra factor that depends on the coupling $\xi$~, ultimately related to the scalar field via \eqref{constraint2}. 
Therefore, there exist three tensionless\footnote{In 2+1 dimensions, a domain wall has a two-dimensional world-volume and it can be viewed as a cosmic string. The action of a domain wall \cite{Vilenkin:1984ib} in two dimensions corresponds to the Nambu--Goto action if one identifies the energy density of the wall with the tension of the string. Therefore, the terms energy density and tension can be used interchangeably.} points at which $\mu$ vanishes, at $x_0 = \pm1$~ and at $x_0 = -\frac{1}{\xi}$~.

We shall use this result to construct accelerating hairy black hole solutions with a single domain wall located at the arbitrary location $x_{\rm brane}\in(x_{\rm min},x_{\rm max})$~, and of which the value is different from $x_0$~, generating the tension
\begin{align}
    \mu = \frac{2A}{\kappa}\sqrt{G(x_{\rm brane})}~,
\end{align}
whose sign can be either positive or negative depending on the sign of the acceleration. Following the terminology of the four-dimensional C-metric \cite{Dias:2002mi}, if $\mu < 0$ the solution is dubbed accelerating hairy black hole pulled by a wall, and if $\mu > 0$ the solution is dubbed accelerating hairy black hole pushed by a strut. 

\subsection{Explicit asymptotically locally anti-de Sitter solutions: Geometric construction and causal structures}\label{SubSec:AdSsols}

From the different families and their corresponding ranges for the transverse coordinate (\autoref{table2}), and from the possible locations of the domain wall topological defect, it is evident that the full spectrum of solutions contained in \eqref{SOL} is very extensive. 
We again expand this
spectrum by considering the three possible cases spanned by the cosmological constant. Under this light, we will construct representative examples primarily for AlAdS geometries, although later asymptotically locally dS and asymptotically locally flat cases will be briefly discussed. 
In addition, when relevant, we will distinguish between slowly and rapidly accelerating black hole phases as well as accelerating particles. Notwithstanding this, it is beyond the scope of this paper to analyze every possible geometry contained in \eqref{SOL}, and so some of the geometries will only be briefly mentioned. Most of the analysis will rely on the description of the most characteristic spacetimes. 

\subsubsection{$\sigma=-1$: Generalizing vacuum class II: Hairy accelerating BTZ black holes}

We commence with the construction of AlAdS solutions. This feature is evidenced by the behavior of the Riemann tensor close to the conformal boundary $\mathcal{I}$~, i.e. 
\begin{equation}
\lim_{x+y\rightarrow0}R^{\mu\nu}{}_{\lambda\rho}\sim-\frac{1}{\ell^2} \delta^{\mu\nu}{}_{\lambda\rho}~,
\end{equation}
where $\ell\in\mathbb{R}$~. As has been previously mentioned, the family of solutions characterized by $\sigma=-1$ naturally connects, in the vacuum limit $\xi=0$~, with the two-parameter extension of the BTZ black hole that is identified as an accelerating BTZ geometry. We therefore start by considering the hairy generalization of the class II solutions described in \autoref{table1}.  
Among the plethora of cases contained in \autoref{table2}, for the sake of simplicity, and to make the construction easier to analyze, we consider the case in which both factors of the polynomial $G(x)$ are positive\footnote{The same convention will be taken for the remaining spacetimes constructed in this paper.}. Therefore, we shall work with the transverse coordinate restricted as follows 
\begin{align}
  0<\xi<1&:  x\in\left(-\frac{1}{\xi},-1\right) \hspace{0.2cm}{\rm or}\hspace{0.2cm} x\in(1,\infty)~,\\
  \xi>1&: x\in(1,\infty)~.
\end{align}
Knowing these restrictions, it is natural to proceed with the accounting of the possible Killing horizons that the geometries might possess. This allows us to properly classify a desired interval $(x_{\rm min}, x_{\rm max})$ on the basis of how many of the Killing horizons are desired on the geometry under consideration. 
As shown in  \autoref{App:KillingHorizons} the most general case exhibits three Killing horizons for the metric function $F(y)$~. For both intervals of $\xi$~, a horizon emerges that does always satisfy the condition $y_{h_3}<-1$~. 
The remaining two horizons, $y_{h_2}$ and $y_{h_1}$~,  will behave accordingly,  with $1<y_{h_2}<y_{h_1}<\frac{1}{\xi}$ if $0<\xi<1$ and $0<y_{h_2}<y_{h_1}<1$ if $\xi>1$~.

Deviations from this general scenario occur under two circumstances. The first is when the acceleration is small ($A<<1$) and only $y_{h_1}$ prevails. The second is when the parameters $A$~, $\ell$ and $\xi$ combine such that $y_{h_2}$ and $y_{h_3}$ merge into a single degenerate horizon. This is possible when $-1/A^2\ell^2$ coincides with the local minimum of the function $(1-y^2)(1-\xi y)$; details are in  \autoref{App:KillingHorizons}.

\begin{figure}[H]
  \begin{minipage}[t]{0.45\linewidth}
    \centering
    \includegraphics[scale=1.5]{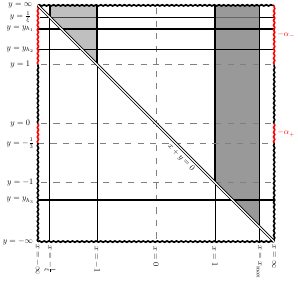}
    \caption{Initial causal structure of the $\sigma=-1$ family for $0<\xi<1$~. The case with three Killing horizons is initially presented. The two allowed domains for the transverse coordinate are depicted in grey. The two branches $\alpha_\pm$ are displayed in red on the vertical axis. They indicate the range of values of $\alpha$ for which the scalar field profile develops a pole.}
    \label{d1}
  \end{minipage}
  \hfill
  \begin{minipage}[t]{0.45\linewidth}
    \centering
   \includegraphics[scale =1.5]{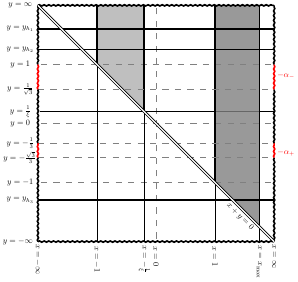}
    \caption{Initial causal structure of the $\sigma=-1$ family for $\xi>1$~. The three Killing horizons case is depicted. The transverse coordinate domain is represented by the grey area. The two branches $\alpha_\pm$ are displayed in red on the vertical axis. They indicate the range of values of $\alpha$ for which the scalar field profile develops a pole.}
    \label{d2}
  \end{minipage}
\end{figure}

To construct interesting geometries we need to select proper sets $(x_{0},x_{brane})$ contained in the ranges of the transverse coordinate. This is needed to proceed with the eventual mirroring and gluing of the spacetime, a step that finally enables us to construct our geometries. In addition, we need to restrict our coordinates in such a manner that the scalar field profile remains everywhere real. 
In the case with three Killing horizons we found the preliminary causal structures described in \autoref{d1} and \autoref{d2}. 

In these figures, we have also included the location of the pole of the scalar field profile. The $y$-coordinate should therefore be constrained in such a way that the reality of $\phi(y,x)$ is everywhere guaranteed. 

With this information at hand, it is customary to define a set $(x_{0},x_{brane})$  within the allowed range of the transverse coordinate. This is chosen on the basis of the eventual Killing horizon structure we desire to have in a given geometry. In order to illustrate our procedure let us consider \autoref{d1} and the interval $(-\frac{1}{\xi},-1)$ of the $x-$coordinate. Generically only the Killing horizons $y_{h_1}$ and $y_{h_2}$ form part of the geometry as the Killing horizon $y_{h_3}$ does not form part of the physically sensible set of coordinates. Next, we consider the $\alpha_+$ branch, ensuring that in the region of interest, the scalar field profile remains everywhere real. 
Both horizons $y_{h_1}$ and $y_{h_2}$ reach conformal infinity (the diagonal double line in \autoref{d1}), and consequently both are non-compact. 

In order to have a physically realistic solution,  we use the freedom encoded in the definition of $(x_{\rm min},x_{\rm max})$~. The following possibilities are at hand. First, we choose $x_{\rm min}$ so that it lies far enough to the right of $-\frac{1}{\xi}$   so that the Killing horizon $y_{h_1}$   does not reach conformal infinity $\mathcal{I}$~. The eventual geometry might then possess one compact and one non-compact (acceleration) horizon. This geometry will represent a hairy rapidly accelerating black hole. A second option is for $x_{\rm min}$ to be chosen further to the right of $-\frac{1}{\xi}$ in such a way that both horizons $y_{h_1}$ and $y_{h_2}$ are not in contact with the conformal boundary. The eventual geometry might possess inner and event horizons. 
It will represent a hairy slowly accelerating extension of the BTZ black hole in which both inner and outer event horizons are present. This behavior finds a cousin in the case of the charged four-dimensional C-metric. The inner horizon in four dimensions that arises in the slowly accelerating case, appears in the three-dimensional setting due to the presence of the scalar field. The scalar field induces in the metric a term that resembles the physics of conformal Maxwell electrodynamics. This due to its conformal invariance.

\begin{figure}[H]
    \centering
    \includegraphics[scale=1]{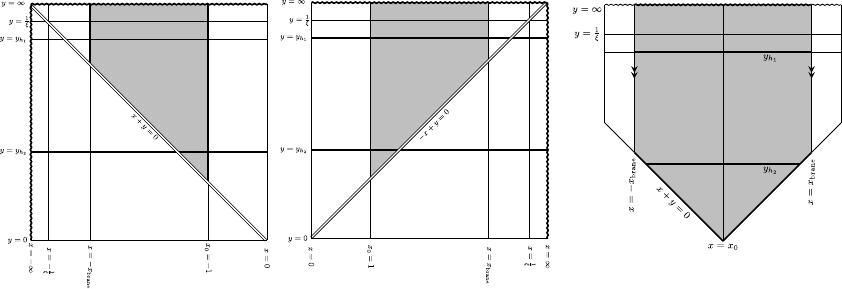}
    \caption{AlAdS rapidly accelerating hairy black hole: This geometry represents a three-dimensional accelerating black hole with both, event and Rindler horizons. It belongs to the family $\sigma=-1$~, where the hair parameter satisfies $0<\xi<1$~. It is supported by a single domain wall, whose position has been achieved by moving the locus $x_{\rm min}$ towards its right hand side, $x_{\rm brane}>-y_{h_1}$~. In this case $x_{\rm max}$ remains as the tensionless locus $x_0$~.}
    \label{fig:side_by_side4}
\end{figure}

\begin{figure}[H]
    \centering
    \includegraphics[scale=1]{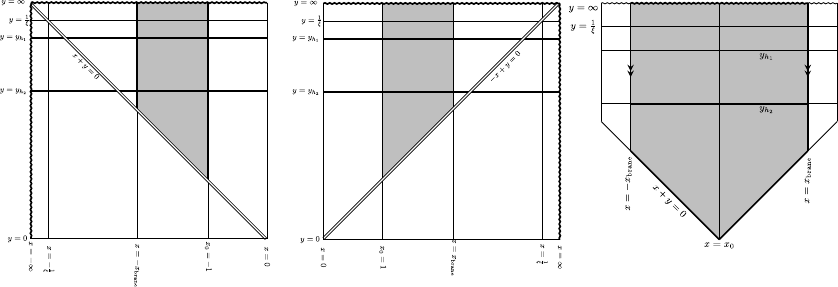}

    \caption{AlAdS slowly accelerating hairy black hole: This geometry represents a three-dimensional accelerating black hole with both, inner and event horizons. It belongs to the family $\sigma=-1$~, where the hair parameter satisfies $0<\xi<1$~. It is supported by a single domain wall, whose position has been achieved by moving the locus $x_{\rm min}$ towards its right hand side, even further than in the previous case, such that no Rindler horizon emerges. $x_{\rm brane}>-y_{h_2}$ and $x_{\rm max}$ remains as the tensionless locus $x_0$~.}  
    \label{fig:side_by_side5}
\end{figure}

Recall that the final spacetime is the one obtained by mirroring these geometries and gluing them via the inclusion of one or more domain walls. In these two cases, only one domain wall is necessary. In fact, as we have moved the value of $x_{\rm min}$ away from $-\frac{1}{\xi}$~, which is a root of the polynomial $G(x)$~, it is possible to use this locus to include a domain wall with a non-trivial tension and produce the gluing. On the same lines, $x_{\rm max}$ has not been modified, and still lies at a value representing a root of $G(x)$~. This point is therefore devoid of any tension and its identification during the mirroring does not produce any topological defect. As previously stated, the location of the domain wall and of the tensionless locus will be always denoted as $x_{\rm brane}$ and $x_0$ and should not be confused with $x_{\rm min}$ and/or $x_{\rm max}$ although they might originally coincide. The complete construction of these solutions, including the corresponding mirroring and gluing, is depicted in \autoref{fig:side_by_side4} and \autoref{fig:side_by_side5}.

Although we do not provide their explicit construction, other solutions can be extracted from \autoref{d1} and \autoref{d2}. Still in the interval $(-\frac{1}{\xi},-1)$ we observe that, keeping fixed $x_{\rm min}$ at $-\frac{1}{\xi}$ but moving $x_{\rm max}$ sufficiently left of $x=-1$~, it is possible to avoid the presence of the non-compact horizon $y_{h_2}$~. This yields the spacetime of an accelerating particle in AdS characterized by the Rindler horizon $y_{h_1}$~. The particle accelerates due to the presence of a domain wall at $x_{\rm brane}=x_{\rm max}$~, while $x_0=x_{\rm min}$ remains as a tensionless locus. A more exotic solution can be found if both $x_{\rm min}$ and $x_{\rm max}$ are moved from their original positions toward the center of the interval. This is performed in order that the horizon $y_{h_1}$ does not reach the conformal boundary and to completely eliminate the presence of the accelerating horizon $y_{h_2}$~. Contrary to the previous examples, here two domain walls are required as both $x_{\rm min}$ and $x_{\rm max}$ are located outside of any root of the polynomial $G(x)$~. This geometry represents a slowly accelerating hairy black hole with a single event horizon. 

Next, we observe that \autoref{d1} and \autoref{d2} share a case in which the transverse coordinate satisfies $x>1$; the resultant geometry thus 
does not depend on the restriction imposed on the hair parameter $\xi$~. In principle, this geometry supports the appearance of three Killing horizons, as now the Killing horizon $y_{h_3}$ belongs to the physical domain of the transverse coordinate.

However the cases $0<\xi\leq1$ and $\xi>1$ each need to be independently analyzed, as they restrict the coordinate $y$ in different ways depending on the values contained in the $\alpha_\pm$ branches. For instance, the negative branch $\alpha_-$ is very restrictive and basically renders the geometry ill-defined, as the scalar field might be imaginary near the region in which the horizons $y_{h_2}$ and $y_{h_1}$ appear. This occurs for any $\xi>0$~. 
On the other hand, values of $\alpha$ contained in $\alpha_+$ allow for a physical solution.
To construct the solution we restrict the value of $x_{\rm max}$~, moving it leftward so that the non-compact horizon is avoided. In other words, the brane is localized so that the horizon $y_{h_3}$ does not form part of the spacetime. In addition, we impose the maximum value of the $y$ coordinate to satisfy $y_{\rm max}>0$ such that the scalar field profile is everywhere real. A hairy slowly accelerating black hole with inner and event horizons is therefore obtained, and its causal structure coincides with the one previously described in \autoref{fig:side_by_side5}. Although in principle $y_{h_3}$ could be included on the spacetime, this situation will always yield a pole in the scalar field, since 
$y_{h_3}<-1$~, violating the condition $y>-\alpha_+$; this holds no matter how small the  $\alpha_+$ branch is.

Prolate coordinates have so far been useful for the integration of the field equations and the subsequent geometric construction of our spacetimes. In order to understand how these new solutions generalize previously known vacuum and non-accelerating spacetimes, it is convenient to switch to polar-like coordinates. This allows for proper identification of a hierarchy of solutions spanned by some of our hairy accelerating geometries. Let us focus on the case in which our geometry describes a hairy accelerating black hole with event and Rindler horizons, \autoref{fig:side_by_side4}. Polar coordinates can be achieved via the transformation 
\begin{align}\label{rescaling}
    y = \frac{m}{Ar}~,\qquad t = A m\tau~,\qquad x = - \cos(m\theta\sqrt{\sigma})~.
\end{align}
 Here the parameter $m$ has been introduced; the aim is to define the proper range of the angular coordinate $\theta$  that will determine which particular geometry is under consideration. Its value is related to the localization of the domain wall $x_{\rm brane}$ so that
\begin{align}
    x_{\rm brane} = -\cos(m\pi\sqrt{\sigma})~,
\end{align}
in turn implying that the compact coordinate always has the range  $\theta \in (-\pi,\pi)$. The domain wall is therefore now located at these identified endpoints.
This identification allows to translate any of the solutions originally constructed in prolate coordinates to a polar-like coordinate domain satisfying $\theta \in (-\pi,\pi)$. 
Complementing this coordinate transformation with the parameter redefinitions 
\begin{align}
    A = {\cal A}m~,\qquad \xi = \zeta{\cal A}~,\qquad \alpha = -\frac{S}{\cal A}~,
\end{align}
yields the solution in the intuitive form 
\begin{align}\label{polarcoords}
    ds^2=\frac{1}{\Omega^2}\left(-F(r)d\tau^2 + \frac{dr^2}{F(r)} + r^2\frac{d\theta^2}{G(\theta)}\right)~,
\end{align}
where the corresponding metric functions and scalar profile are given by 
\begin{align}
    \Omega ={}& 1-\mathcal{A} r \cosh (m\theta )~,\nonumber\\
    F(r) ={}& -\Lambda r^2 -m^2\left(1-{\cal A}^2r^2\right)\left(1-\frac{\zeta}{r}\right)~,\nonumber \\ G(\theta) ={}& 1-{\cal A}\zeta\cosh(m\theta)~,\nonumber \\  \phi(r,\theta) ={}& \sqrt{\frac{8}{\kappa}}\sqrt{\frac{1-\mathcal{A} r \cosh (m\theta )}{1-r S}}~.
\end{align}
and where we have already selected $\sigma=-1$ in order to connect with the solution under discussion.
Notice that in these coordinates the Kretschmann scalar 
\begin{align}
   {\mathsf{K}} =12 \Lambda ^2 + \frac{6m^4 \zeta ^2 \left(\mathcal{A} r \cos \left(\theta  m \sqrt{\sigma }\right)-1\right)^6}{r^6}~,
\end{align}
diverges at $r=0$~, highlighting the previously identified curvature singularity at $y\rightarrow\infty$;   conformal infinity is now located at $r = \left[\cA \cos\left(m\theta\sqrt{\sigma}\right)\right]^{-1}$~.

The main advantage of these coordinates relies on their use to determine the subcases in which the acceleration parameter vanishes and/or when the scalar field backreaction is null. Notice at this point that both limits are independent. We commence by considering the cases in which the scalar field backreaction vanishes. For this to be case we need to seek  values of the parameters such that $\zeta$ goes to zero. 
In these coordinates we find that \eqref{constraint1} and \eqref{constraint2} become
\begin{align}
    \zeta = \frac{2 S}{3 S^2-\mathcal{A}^2}~, \qquad \lambda = -\frac{\kappa^2}{512\ell^2}\left(\frac{\cA^2-3S^2+m^2\left(S^2-\cA^2\right)^2\ell^2}{3S^2-\cA^2}\right)~.
\end{align}
It can be noticed that the hair parameter $\zeta$ vanishes by means of two independent limits, i.e $S\rightarrow\infty$ or $S\rightarrow0$~. In the first case  
\begin{align}
    \lim_{S\to\infty} \phi(r,\theta) = \lim_{\zeta\to 0} \phi(r,\theta)= 0~,
\end{align}
and thus no scalar field configuration remains, and the metric functions recover their vacuum form.
Notice, however, that this limit is subtle as the self-interaction coupling  $\lambda\to\infty$~. Nonetheless, the on-shell action \eqref{action1} can be proven to be well-defined,  recovering  the form of the Einstein--Hilbert action 
\begin{align}
    \lim_{S\to \infty} I[g_{\mu\nu},\phi] = \int_{\cal M}d^3x \sqrt{-g}\left(\frac{R-2\Lambda}{2\kappa}\right)~,
\end{align}
from which the vacuum solution emerges.
This makes the limit in which the scalar field vanishes well-defined, not only at the level of the metric and scalar configuration, but at the level of the action principle as well. Consequently  our hairy rapidly accelerating black hole is connected in the $S\to\infty$ limit with the rapidly accelerating extension of the BTZ black hole. This solution will henceforth dubbed  the hairy accelerating BTZ solution.

On the other hand, we are entitled to take the limit $S\rightarrow0$~. In this case, the metric functions remain the ones of the accelerating vacuum geometries but with a non-trivial scalar field  
\begin{align}\label{stealthphi}
    \lim_{S\to 0} \phi(r,\theta) = \lim_{\zeta \to 0}\phi(r,\theta) = \sqrt{\frac{8\Omega}{\kappa}}~,
\end{align}
that  does not backreact on the geometric structure of the spacetime metric.  For this to be the case we must also have
\begin{align}
   \lim_{S\to 0}\lambda = \frac{\kappa}{512\ell^2}\left(1+m^2\cA^2\ell^2\right)~.
\end{align}
These types of solutions are known as stealth configurations and are characterized by a non-trivial matter source that yields a solution of the geometric and matter sectors of Einstein equations independently: $G_{\mu\nu}=0=T_{\mu\nu}$~. We dub this configuration the stealth accelerating BTZ solution. 

Finally, let us disclose what occurs when the non-accelerating limit is performed. 
The null acceleration limit provides us with static spherically symmetric hairy configurations, namely, the Henneaux--Martinez--Troncoso--Zanelli (HMTZ) \cite{Henneaux:2002wm} and the Martinez--Zanelli (MZ) \cite{Martinez:1996gn} black holes. 
Setting $\cA=0$ and using the redefinitions 
\begin{align}
    \zeta = -\frac{2B}{3}~,\qquad m^2 = \frac{3B^2}{\ell^2}(1+\nu)~,\qquad S = -\frac{1}{B}~, 
\end{align}
our configuration becomes 
\begin{align}\label{hairylimit}
    ds^2={}&-F(r)dt^2+\frac{dr^2}{F(r)}+r^2d\theta^2~,\nonumber \\ F(r) ={}& \frac{r^2}{\ell^2} - (1+\nu)\left(\frac{3B^2}{\ell^2}+\frac{2B^3}{\ell^2 r}\right)~, \nonumber \\
    \phi(r) ={}& \sqrt{\frac{8}{\kappa}}\sqrt{\frac{B}{B+r}}~, 
\end{align}
where $\lambda=-\frac{\kappa^2\nu}{512 \ell^2}$~. This corresponds to the HMTZ black hole and the limit of vanishing $\nu$ provides the original non-self-interacting solution known as the MZ geometry. 
We note that there is no limit from where this hairy BTZ extension recovers the standard BTZ black hole. Indeed, 
the mass parameter $B$ appears in the scalar field, which vanishes only for $B=0$~. Hence the absence of the scalar field yields  massless BTZ solution. 

\begin{figure}[h!]
    \centering
\includegraphics{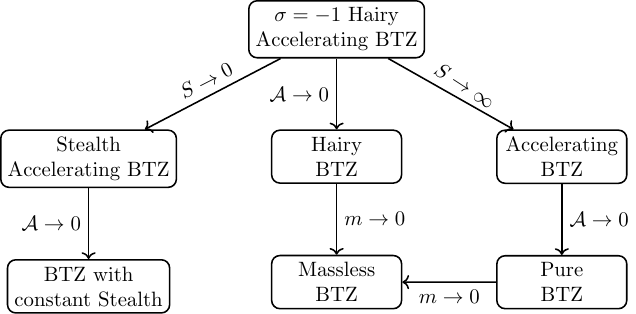}
    \caption{Hierarchy of solutions contained int $\sigma = -1$ hairy accelerated black hole.}
    \label{fig:hierarchiessm1}
\end{figure}

In summary, the hairy accelerating BTZ black hole is shown to be connected with three classes of geometries:  the accelerating BTZ black hole, the accelerating BTZ black hole dressed by a stealth scalar field, and the known family of hairy spherically symmetric solutions dubbed the HMTZ and MZ black holes. 
The whole hierarchy is depicted in 
\autoref{fig:hierarchiessm1}~.
It is interesting to notice that, due to the presence of the acceleration parameter, it is possible to connect the hairy configuration with the standard vacuum BTZ black hole. This is possible via the $S\rightarrow\infty$ limit, and it represents a novel limit usually not present in non-accelerating hairy configurations, not only in dimension three, but in dimension four as well.  

\subsubsection{$\sigma=1$: Generalizing vacuum class I: Hairy accelerating black holes with no BTZ limit}

The family of solutions given by $\sigma=1$ is, on the other hand, known to connect with the class I  solutions listed in \autoref{table1}. This vacuum class primarily represents accelerating particles in AdS with the exception of a geometry, dubbed Ic, that represents an accelerating extension of the BTZ black hole with no continuous limit to the standard BTZ geometry. Here we explicitly construct the hairy generalization of this class, with our main focus on the   extensions that represent black holes. The ranges of the transverse coordinates in this case are given  by 
\begin{align}
  0<\xi<1&:  x\in(-1,1)~,\\
  \xi>1&: x\in\left(-\frac{1}{\xi},1\right).
\end{align}
The structure of the Killing horizons is discussed in  \autoref{App:KillingHorizons}. For both restrictions on $\xi$ the Killing horizon $y_{h_3}$ always belongs to the domain $(-1,0)$~, while $y_{h_2}$ and $y_{h_1}$ satisfy either $y_{h_2}<1<\frac{1}{\xi}<y_{h_1}$ for $0<\xi<1$ or $y_{h_2}<\frac{1}{\xi}<1<y_{h_1}$ for $\xi>1$~, unless the numerical value of the cosmological constant is so large that $y_{h_2}$ becomes negative. This   stage is prior to the merging of $y_{h_2}$ and $y_{h_3}$ into the extremal case.
Generically speaking both cases express the same geometrical properties. The Killing horizons $y_{h_3}$ and $y_{h_2}$ are non-compact --  they reach conformal infinity, whereas $y_{h_1}$ does not. 

\begin{figure}[H]
    \centering
\includegraphics[scale=1.2]{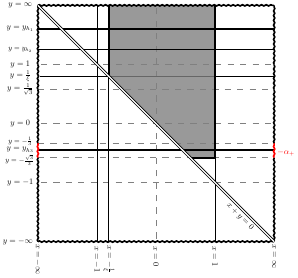}
    \caption{Initial causal structure of the $\sigma = 1$ family for $\xi>1$~. Only the relevant branch $\alpha_+$ is displayed. The allowed domain of the transverse coordinate is depicted in grey.}
    \label{fig:causal}
\end{figure}

Physically meaningful solutions can be constructed by introducing a single domain wall. We focus on the case with $\xi>1$ as it allows for a rich horizon structure whilst maintaining a real scalar field profile. The preliminary causal structure of the solution is depicted in \autoref{fig:causal} where, as in \autoref{d1} and \autoref{d2}, we have located the interval $(x_{\rm min},x_{\rm max})$~, the Killing horizons, and the corresponding branch $\alpha_+$ we are considering. 

\begin{figure}[h]
   \centering
   \includegraphics[scale=1]{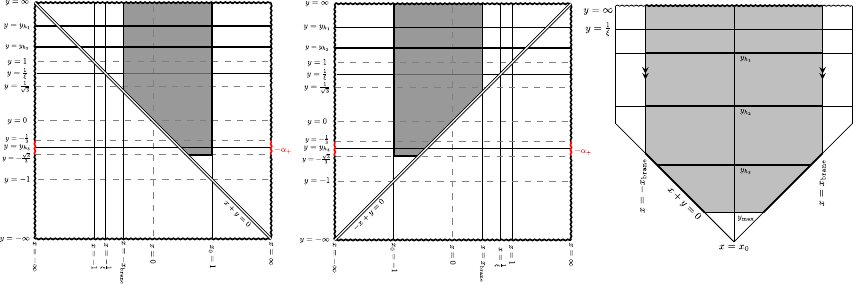}
    \caption{AlAdS rapidly accelerating hairy black hole with inner and event horizon: This geometry represents a three-dimensional accelerating black hole with inner, event, and Rindler horizons. It belongs to the family $\sigma=1$~, where the hair parameter satisfies $\xi>1$~. It is supported by a single domain wall, whose position has been achieved by locating $x_{\rm max}$ at a finite locus satisfying $x_{\rm brane}>-y_{h_3}$~. $x_{\rm max}$ remains as the tensionless locus $x_0$~.}
    \label{fig:side_by_side7}
\end{figure}

\begin{figure}[h]
    \centering
    \includegraphics[scale=1]{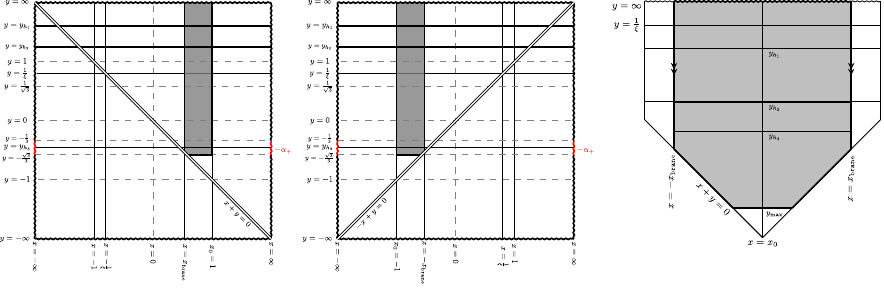}
    \caption{ AlAdS slowly accelerating hairy black hole with three Killing horizons: This geometry represents a three-dimensional accelerating black hole with three compact horizons. It is supported by a single domain wall located at $x_{\rm brane}>-y_{h_3}$ such that no horizon reaches conformal infinity. It belongs to the family $\sigma=1$; here $\xi>0$~.}
    \label{fig:side_by_side8}
\end{figure}

Two relevant spacetimes can be constructed. 
In one, $x_{\rm brane}$ is located sufficiently rightward of $x_{\rm min}$ to ensure that the Killing horizon does not reach conformal infinity. This gives rise to a solution featuring inner and event horizons, supplemented by one accelerating horizon. We depict this geometry in
\autoref{fig:side_by_side7}.
Alternatively,  the location of the brane can be such that even the horizon $y_{h_3}$ does not reach conformal infinity. In this case the geometry becomes slowly accelerating and features three compact horizons. This seems to be a new class of multi-horizon accelerating spacetimes,  which we illustrate in  \autoref{fig:side_by_side8}. In addition, a black hole with event and Rindler horizons and a slowly accelerating black hole with a single event horizon are obtained by appropriately locating the brane on the left side of $x_{\rm max}$~. 

Recall that, for all solutions to be physical, proper values of $\alpha$ need to be selected in order for the scalar field profile to remain real. In the cases of the geometries depicted in \autoref{fig:side_by_side7} and \autoref{fig:side_by_side8} the positive branch $\alpha_+$ must be selected and it must be properly restricted such that $y_{h_3}>y_{\rm max}>-\alpha_+$~. 
This is the case for large values of the cosmological constant. There, the value of $y_{h_3}$ tends to approach $y=0$ from below. $\alpha_+$ can therefore be restricted so that  $y_{h_3}>y_{\rm max}>-\alpha_+$ is satisfied before $y_{h_2}$ becomes negative.  
We can then choose a convenient value of $y_{\rm max}$ so that the scalar field has no poles.

We note that in the AlAdS cases described  $\sigma=-1$ no geometries with three horizons can be achieved. This is because $y_{h_3}<-1$ implying that all branches $\alpha_{\pm}$ render the solution ill-behaved if the region containing $y_{h_3}$ is considered as part of the spacetime. 
\\

\begin{figure}[h!]
    \centering
\includegraphics{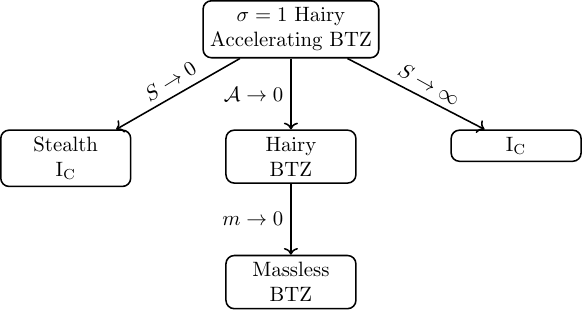}
    \caption{Hierarchy of solutions contained int $\sigma = 1$ hairy accelerated black hole.}
    \label{fig:hierarchiessp1}
\end{figure}

Moving to polar-like coordinates the metric and scalar profile read 
\begin{align}
    \Omega ={}& \mathcal{A} r \cos (m\theta )-1~,\nonumber\\
    F(r) ={}& -\Lambda r^2 -m^2\left({\cal A}^2r^2-1\right)\left(1-\frac{\zeta}{r}\right)~,\nonumber \\ G(\theta) ={}& 1-{\cal A}\zeta\cos(m\theta)~,\nonumber \\  \phi(r,\theta) ={}& \sqrt{\frac{8}{\kappa}}\sqrt{\frac{\mathcal{A} r \cos (m\theta )-1}{r S-1}}~. 
\end{align}
As  with the previous family ($\sigma=-1$), polar coordinates allow for transparent analysis of the vacuum and non-accelerating limits. Again,  $\zeta=0$ can be achieved by means of the two limits $S\rightarrow\infty$ or $S\rightarrow0$~. To show the corresponding limits,   consider the case in which the causal structure of the $\sigma=-1$ family describes a rapidly accelerating hairy black hole. The limit $S\rightarrow\infty$ connects with the vacuum class I. This class, depending on the value of the parameter $m$ will describe either accelerating particles in AdS or an accelerating extension of the BTZ black hole with no limit to the standard BTZ geometry. This latter spacetime is known as class Ic. On the other hand, the limit $S\rightarrow0$ brings us to a stealth generalization of the classes I and Ic, while the non-accelerating limit $A\rightarrow0$ yields the hairy spherically symmetric configurations represented by the HMTZ spacetime. Observe that, as there is no limit between the Ic class and the BTZ black hole, our family $\sigma=-1$ does not reach the BTZ standard geometry in any of its possible limiting cases. This is expressed graphically in \autoref{fig:hierarchiessp1}~.

\section{Comments on Holography: Regularized action, holographic stress tensor, and spacetime mass}\label{Sec:holography}

A striking characteristic of accelerating AlAdS solutions is that they feature a non-trivial conformal boundary, non-trivial in the sense that it is located at the surface 
$\Omega(r,\theta)=0$~, instead of corresponding to a constant value of the usual holographic radial coordinate. 
Consequently, the embedding of accelerating solutions in the standard holographic picture is very cumbersome, as the definition of a holographic coordinate becomes less intuitive. In \cite{Arenas-Henriquez:2022www}, three-dimensional black holes have been studied in the framework of the Fefferman--Graham gauge, in which the metric can be written as 
\begin{align}\label{FG}
    ds^2 = \frac{\ell^2}{z^2}dz^2 + h_{ij}dx^i dx^j~.
\end{align}
Here, the boundary metric admits a near-to-boundary expansion of the form
\begin{align}
    h_{ij} = \frac{\ell^2}{z^2}\left(g_{(0)ij} + z^2\left( g_{(2)ij} + h_{(2)ij} \log z\right)+\dots\right).
 \end{align}
To achieve this one can apply a coordinate transformation \cite{Anabalon:2018ydc},  
 \begin{eqnarray}\label{coordFG}
y&=&\cos\rho+\sum\limits_{m=1}^{\infty} F_{m}(\rho)z^{m}~, \nonumber \\
x&=& -\cos \rho +\sum\limits_{m=1}^{\infty}G_{m}(\rho)z^{m}~,
\end{eqnarray}
such that, the new coordinate $\rho$ is now perpendicular to the (conformal) boundary of the spacetime. The functions $F_m(\rho)$ and $G_m(\rho)$ are fixed by requiring the asymptotic form \eqref{FG}, with no cross terms of the form $g_{z i}$~. All coefficients in the expansions can be solved order by order, with the exception of $F_1(\rho)$~, 
which cannot be fixed; it appears as a conformal factor of the boundary metric $g_{(0)}$~. This is consistent with the fact that the conformal boundary of AdS does not correspond to a fixed metric but to a conformal equivalence class.
Notwithstanding, it has been recently shown   \cite{Arenas-Henriquez:2023hur} that this method fails to reproduce holographic quantities in three dimensions, as the stress tensor of a two-dimensional CFT is a quasi-primary, transforming non-trivially under conformal transformations. Recall that the Weyl anomaly is a conformal invariant, and therefore does not dependent on the conformal representative and  can be still obtained through this method.

An alternative 
to this procedure makes use of 
more intuitive coordinates to describe the boundary \cite{Hubeny:2007xt,Cassani:2021dwa,Arenas-Henriquez:2023hur}
\begin{align}\label{Hologzpolar}
    z:=\frac{1}{r} - {\cal A}\cos(m\theta\sqrt{\sigma})~,
\end{align}
 or in its prolate fashion
\begin{align}\label{Hologzcanonical}
    y = z - x~. 
\end{align}
The conformal boundary is now at the locus $z=0$~, instead of  the  surface $\Omega=0$
(where $r$ is not constant). The  price   paid for this change is that   the boundary metric now features a non-diagonal term, and so the matching with \eqref{FG} is only valid at the leading order. Nonetheless, this is enough to obtain the holographic stress tensor, as only the leading order contributes. Moreover, the extrinsic curvature has the correct fall-off at   leading order
\begin{align}
    K^i_j \sim \frac{1}{\ell}\delta^i_j + \dots~,
\end{align}
which suffices to compute holographic quantities \cite{Miskovic:2006tm}. Hence, we can apply this strategy to the case of the hairy accelerating black holes described in the previous sections. 

First, note that the  action \eqref{action1}
can be written on-shell as 
\begin{align}
    I[g_{\mu\nu},\phi] = -\frac{1}{2}\int_{M}d^{d+1} x\sqrt{-g}\left(\left(\partial_\mu \phi\right)^2+{\tilde m}^2\phi^2 + \lambda\phi^6+\frac{4}{\ell^2\kappa}\right)~,
\end{align}
where $d=2$~, and  $\tilde{m}^2:=-\frac{3}{4\ell^2}$~. This is the action of a massive scalar field on an AlAdS background featuring a hexic self-interacting potential. To cast the action in this form we have used the fact that $R=6\Lambda$~, as dictated by the bulk conformal symmetry of the matter sector. The mass $\tilde{m}$ is indeed admissible within the unitary Breitenlohner--Freedman bound \cite{Breitenlohner:1982bm,Witten:1998qj,Klebanov:1999tb}
\begin{align}
   -1 < {\tilde m}^2\ell^2 < 0~. 
\end{align}
Now, let us proceed by using the holographic coordinate \eqref{Hologzcanonical} on our family of solutions characterized by $\sigma=-1$~. The scalar field near the boundary ($z=0$) behaves as 
\begin{align}
    \phi \sim z^{\Delta_-}(\phi_-(x) + \dots)+z^{\Delta_+}(\phi_+(x)+\dots)~,
\end{align}
where $\Delta_\pm = 1\pm \frac12$ are the two solutions of
\begin{align}
    \Delta(\Delta-2)={\tilde m}^2\ell^2~,
\end{align}
representing the conformal dimension of the boundary operator, and 
\begin{align}
    \phi_\pm(x) = \mp \sqrt{\frac{5\mp 3}{\kappa}}\left(\alpha - x\right)^{-\Delta_\pm}~.
\end{align}
Unitarity of the dual theory implies that $x < \alpha$~, which is consistent with the fact that we are considering regions of the spacetime where the scalar field is real and regular. 
The expansion can be alternatively written as
\begin{align}
    \phi(z,x) = z^{d-\Delta_\pm}\left(\phi_0(x) + \dots\right) + z^{\Delta_\pm}\left(\phi_1(x)+\dots\right)~,
\end{align}
where $\phi_0(x)$   corresponds to the source of the boundary operator, and $\phi_1(x)$  describes physical fluctuations determined by $\phi_0(x)$~. 
This implies different quantum theories at the boundary, regardless of
 the choice of 
 $\Delta_+$ or $\Delta_-$ as the conformal weight of the dual operator. 

However  in order to have well-defined boundary conditions at the boundary, it is necessary to keep  the renormalized momentum fixed \cite{Papadimitriou:2007sj}
\begin{align}
    {\hat \pi}_{(\Delta_+)} = \lim_{z\to 0}z^{-\Delta_+} \pi_{(\Delta_+)}~,
\end{align}
instead of $\phi_+$~. This is given in terms of an expansion of the canonical momenta conjugate to the scalar field 
\begin{align}
    \pi_\phi  =\sqrt{h}\left(\pi_{(\Delta_-)} + \dots + \pi_{(\Delta_+)}+\dots\right)~,
\end{align}
that for the action \eqref{action1} reads
\begin{align}
    \pi_\phi = \sqrt{h}\left( n^\mu\partial_\mu \phi + \frac{1}{8}\phi K \right)~. 
\end{align}
Moving forward, the geometry in which the CFT lies is described by
\begin{align}\label{g0}
    g_{(0)ij} = \lim_{z\to0}z^2h_{ij} = \frac{1}{A^2}\left( -F(x)dt^2 + \frac{dx^2}{F(x)G(x)} \right)~,
\end{align}
where
\begin{align}
    F(x) = \lim_{z\to 0} F(z-x) = (1-x^2)(1+\xi x) + \frac{1}{A^2\ell^2}~,
\end{align}
and $G(x)$ retains its original form, as it does not depend on $y$~. The boundary metric has a non-constant curvature given by 
\begin{align}
    {\cal R}[g_{(0)}] = \lim_{z\to0} z^{-2} R[h]=\frac{1}{2} A^4 \ell ^2 \left(4 \xi  \left(7 x^2-5\right) x+8 x^2+\xi ^2 \left(21 x^4-18 x^2+1\right)-4\right)~.
\end{align}
The curvature of the non-hairy accelerating black hole \cite{Arenas-Henriquez:2023hur} is recovered in the limit of vanishing $\xi$~, and vanishes in the zero acceleration limit. 

Let us now move to the characterization of the holographic stress tensor. 
It has been shown \cite{Arenas-Henriquez:2023hur}   that acceleration in three dimensions produces new divergences in the bulk action, and that the standard holographic renormalization must be supplemented with new terms that are related to the physics of the domain wall. The domain wall introduces a new internal boundary in the spacetime. Consequently, on each side of the wall, the corresponding GHY terms are needed. 

As a result, the action consists of the renormalized action of \cite{Balasubramanian:1999re}, the counterterms related to the presence of the scalar field (see below), and the extra terms that are localized on the domain wall, whose variations with respect to the wall metric provide the Israel junction conditions. Due to the fact that these new terms in the action depend exclusively on the domain wall metric, the covariant structure of the holographic tensor does not suffer modifications with respect to the non-accelerating case. Recall that the domain wall is the reason behind the accelerating nature of the solutions. \\

In order to understand the computation of the holographic mass via the holographic tensor let us first tackle the non-accelerating hairy case \cite{Henneaux:2002wm}, which is given by the $\cA\to0$ limit of the $\sigma=-1$ AlAdS hairy accelerated solution \eqref{hairylimit}. The counterterms that give rise to the renormalized Euclidean action, and thence the holographic stress tensor, have been previously identified for the case of a minimally coupled scalar field \cite{Gegenberg:2003jr}. It reads 
\begin{align}\label{Iphi}
    I_{\phi} = \frac{1}{4}\int_{\partial {\cal M}} d^2 x\sqrt{h}\left(\frac{\phi^2}{2\ell} - \phi {\hat n}^\mu\partial_\mu\phi\right)~,
\end{align}
where ${\hat n}^\mu$ is the outward pointing normal to the conformal boundary. This term, along with the corresponding GHY term and the Balasubramanian–Krauss counterterm for the Einstein gravity sector in the action, define the complete action principle of Einstein gravity supplemented with a minimally coupled scalar field. 
Notice that the second term in \eqref{Iphi} is extrinsic to the boundary; however, it can be replaced by a term intrinsic to the boundary if mixed boundary conditions are taken into account \cite{Anabalon:2015xvl}. 

Moving to the conformal frame, where our theory takes place, the GHY term is indeed modified due to the nontrivial coupling between the Ricci curvature and the scalar field, $\sim R\phi^2$~. However, we have proven that due to the decay of the scalar field near the conformal boundary, the counterterms \eqref{Iphi} make the on-shell action finite in the conformal frame as well.
Therefore the total Euclidean action reads
\begin{align}\label{Iren}
    I_{\rm ren} = -I[g_{\mu\nu},\phi]  {}&+\frac{1}{\kappa}\int_{\partial{\cal M}}d^3 x \sqrt{h}\left[\left(1-\frac{\kappa}{8}\phi^2\right)K - \frac{1}{\ell} + \frac{\kappa}{4}\left(\frac{\phi^2}{2\ell} - \phi {\hat n}^\mu\partial_\mu\phi \right)\right] \nonumber \\ {}&-\log z\int_{\partial\cal M}d^2x\sqrt{h}\langle T^i_{~i}\rangle~,
\end{align}
where the first term is the Euclidean version of the bulk action \eqref{action1}, and the boundary term is proportional to the extrinsic curvature. For completeness, we have also included the counterterm that cures an eventual logarithmic divergence arising for curved backgrounds \cite{Emparan:1999pm}. This divergence can be associated with the Weyl anomaly. Its counterterm does not contribute to the boundary stress tensor in three dimensions as it is related to a topological invariant of the boundary theory \cite{ deHaro:2000vlm}. \\

The evaluation of the action in the hairy black hole of \cite{Henneaux:2002wm} proves itself finite and it 
can be related to the Gibbs free energy $F$ that satisfies the quantum statistical relation of black hole thermodynamics
\begin{align}
    I_{\rm ren} = \beta M - S = \beta F~.
\end{align}
As usual, $\beta$ is the inverse of the temperature, 
\begin{align}
    S = \frac{2\pi}{\kappa}\left(\frac{2 \pi r_h^2}{B+r_h}\right)~,
\end{align}
is the black hole entropy, and
\begin{align}
    M = \frac{3\pi B^2}{\kappa \ell^2}(1+\nu) = \frac{m^2}{8G}~,
\end{align}
is the black hole mass. This mass result can be corroborated by means of the holographic tensor.
The holographic stress tensor associated with \eqref{Iren} is given by
\begin{align}\label{Tij}
    \langle T_{ij}\rangle ={}& \lim_{z\to 0}\left(\pi_{ij} + \frac{2}{\sqrt{-h}}\frac{\delta L_{\rm ct}}{\delta h^{ij}}\right)~,\nonumber \\ ={}&-\frac{1}{\kappa}\lim_{z\to 0}\left(1-\frac{\kappa}{8}\phi^2\right)\left(K_{ij} - K h_{ij} + \frac{1}{\ell}h_{ij}\right)~,
\end{align}
where $L_{\rm ct}$ is the Lagrangian density associated with the counterterms of \eqref{Iren}, and
\begin{align}
    \pi_{ij} = -\frac{1}{\kappa}\left[\left(1-\frac{\kappa}{8}\phi^2\right)\left(K_{ij}-Kh_{ij}\right) + \frac{\kappa}{4}\phi {\hat n}^\mu \partial_\mu \phi\right]~,
\end{align}
is the canonical momentum associated with the radial evolution of the spacetime geometry. Notice that the last term in the canonical momentum cancels with the variation coming from the last term in the counterterm \eqref{Iren}. We get
\begin{align}
    \langle T^i_{~j}\rangle = \frac{3 B^2 (1+\nu)}{2 \kappa  \ell }{\rm diag}(-1,1)~,
\end{align}
which is traceless, and the associated energy
\begin{align}\label{HMTZmass}
    M = \int d\theta \sqrt{g_{(0)}}\langle T^\tau_{~\tau}\rangle = \frac{3 \pi B^2 (1+\nu)}{\kappa  \ell^2 }~,
\end{align}
in agreement with the black hole mass in the Einstein frame obtained via Hamiltonian methods \cite{Henneaux:2002wm, Gegenberg:2003jr}.
The Hamiltonian for computing the mass in the conformal frame yields an extra term with respect to the Einstein frame \cite{Cardenas:2014kaa}. This extra contribution depends on the boundary conditions that the scalar field satisfies and vanishes if they correspond to mixed boundary conditions. Therefore, for the HMTZ solution, the mass in both frames is the same. In what follows we show that the scalar field of the accelerating hairy black hole satisfies mixed boundary conditions, and therefore use of the holographic stress tensor \eqref{Tij}  to ascertain the spacetime mass is warranted.

We proceed now with the computation of the black hole mass for our accelerating hairy black holes. It can be seen from \eqref{Iren}, that under the presence of acceleration, an extra divergence of order ${\cal O}\left(z^{-1}\right)$ takes place. This divergence cannot be removed by previously studied counterterms  \cite{Arenas-Henriquez:2023hur} for accelerating black holes in pure Einstein-$\Lambda$ gravity; in fact, it was proven to cancel only by the contribution of the domain wall to the Euclidean action.  
The domain wall action is constructed such that one obtains the Israel junction equations via the variational principle \cite{Gregory:2001dn, Gregory:2001xu, Charmousis:2006pn, Arias:2019zug, Arenas-Henriquez:2023hur}.

The domain wall contribution to the action is then composed of two terms, the domain wall action \cite{Vilenkin:1984ib}, and the generalized GHY terms on each side of the brane \cite{Aviles:2019xae}, i.e.,
\begin{align}
    I_{\Sigma} = -\frac{1}{\kappa}\int_{\Sigma}d^2x\sqrt{\gamma}\left(\left(1-\frac{\kappa}{8}[\phi]^2\right)[{\cal K}] -\mu\right)~.  \label{brane}
\end{align}
Then, the full action, given by  \eqref{Iren} and \eqref{brane}, turns to be finite when evaluated on the hairy accelerating solutions. The divergence is exactly canceled by the brane contribution to the action. As we have stated before, \eqref{brane} does not modify the covariant structure of the holographic tensor. Therefore we can use \eqref{Tij} in this hairy accelerating case as well, obtaining 
\begin{align}
    \langle T^\tau_{~\tau}\rangle ={}&  \frac{A^2\ell}{8\kappa}\left( A^2\ell^2\left(\xi ^2 \left(33 x^4+1\right)+6 \left(2-5 \xi ^2\right) x^2-4 \xi  \left(11 x^2+9\right) x-8\right)-4-12\xi x \right)~, \nonumber \\ \langle T^x_{~x}\rangle ={}& \frac{A^2\ell}{8\kappa}\left(A^2 \ell ^2 (\xi -x (3 \xi  x+2))^2+4+12 \xi  x\right)~.
\end{align}
This furnishes us with the spacetime mass 
\begin{align}
    M ={}& \int dx \sqrt{g_{(0)}}\langle T^\tau_{~\tau}\rangle \nonumber \\ ={}& \int dx \left(\frac{4+12x-A^2 \ell ^2 \left(\xi ^2+33 \xi ^2 x^4+44 \xi  x^3+6 \left(2-5 \xi ^2\right) x^2-36 \xi  x-8\right)}{8 A \kappa  \sqrt{\left(1-x^2\right) (1+\xi  x)}}\right)~,\label{holomass}
\end{align}
which generically is given in terms of elliptic integrals. Recall that, in order for this computation to make sense, the original geometry must be that of a slowly accelerating black hole. This means, the black hole under consideration exhibits one event horizon and is devoid of any Rindler causal structure obstruction. 
In prolate coordinates, this mass does not possess a well-behaved $A\to 0$ limit. This is an artifact of the coordinates, as polar-like coordinates make this limit transparent. It can be checked that the non-accelerating limit of $\eqref{holomass}$ correctly reproduces \eqref{HMTZmass}, while the vacuum limit reproduces the one found in \cite{Arenas-Henriquez:2023hur}. In addition, the holographic stress tensor displays a Weyl anomaly
\begin{align}
    \langle T^i_{~ i}\rangle = \frac{c}{24\pi}{\cal R}[g_{(0)}]~,
\end{align}
where $c=12 \pi  \ell/\kappa = 3\ell/2G$ equals the Brown--Henneaux central charge \cite{Brown:1986nw}, in agreement with \cite{Henneaux:2002wm, Gegenberg:2003jr, Arenas-Henriquez:2023hur}. This makes manifest the fact that neither the scalar nor the acceleration introduces modifications to the conformal anomaly. This result is consistent with the fact that boundary conditions for the scalar field preserve the conformal symmetry of the boundary theory \cite{Henneaux:2002wm, Henneaux:2004zi}. 

Observe that the holographic stress tensor is not covariantly conserved with respect to the boundary metric, and it gives the Ward identity
\begin{align}
    \nabla_j \langle T^j_{~ i}\rangle = \frac{3A^2\ell\xi}{2\kappa}\delta_i^x~,
\end{align}
which is related to the source and current of the dual scalar operator \cite{Papadimitriou:2004ap}.
Moreover, just as for the case of three-dimensional accelerating black holes in vacuum \cite{Arenas-Henriquez:2023hur}, the dual stress tensor can be written as the one of a perfect fluid on a curved background
\begin{align}
    \langle T_{ij}\rangle  = p \left(u_i u_j + g_{(0)ij}\right) + \rho u_i u_j~.
\end{align}
Here 
\begin{align}
    u^i = \frac{1}{\sqrt{-g_{(0)tt}}}\left(\frac{\partial}{\partial t}\right)^i~, \qquad u^2 = -1~,
\end{align}
is the two-velocity of the fluid, and 
\begin{align}
    \rho = \langle T^t_{~t}\rangle~,\qquad p = \langle T^x_{~x}\rangle~,
\end{align}
corresponding to the energy density and pressure of the fluid, respectively.  

Finally, we   stress that the scalar field satisfies mixed boundary conditions
\begin{align}
    \phi_+(x) = -\ell\gamma\phi^3_-(x)~,
\end{align}
with a finite deformation parameter $\gamma = \frac{\kappa}{16\ell}$~. Then, in order to have a well-posed variational principle, one needs to add an extra term \cite{Papadimitriou:2007sj, deHaro:2006wy, deHaro:2006ymc, Aparicio:2012yq}
\begin{align}\label{multitrace}
    I_0 = -\int_{\partial\cal M}\sqrt{g_{(0)}}f(\phi_-) = -\frac{\ell\gamma}{4}\int_{\partial\cal M} d^2x\sqrt{g_{(0)}} \phi_-^4(x)~,
\end{align}
where $f(\phi_- )$ is a deformation function that depends on the boundary condition of the scalar field. This new term produces a new contribution to the 1-point function of the boundary scalar operator, and it produces a marginal multi-trace deformation in the dual theory \cite{Witten:2001ua, Papadimitriou:2007sj}. 
Moreover, the boundary source that produces connected diagrams is given by \cite{Papadimitriou:2007sj}
\begin{align}\label{J}
    J = -{\hat \pi}_{(\Delta_+)} - f'(\phi_- )~.
\end{align}
For the scalar configuration of the hairy accelerating black hole one gets that ${\hat\pi}_{(\Delta_+)} = -(\Delta_+-\Delta_-)\phi_+/\ell = \gamma \phi_-^{3}$~, 
such that the dual source vanishes, i.e., $J  = 0$~. This is a physical condition that is needed in order to obtain the holographic $n$-point functions\footnote{This is in contrast to single trace deformations, which require a non-zero source, such as the glueball scalar operator.}. 
It would be interesting to understand how to implement this term along the lines of  \cite{Anabalon:2015xvl}, where the extrinsic boundary counterterm $~n^\mu\partial_\mu\phi$ is replaced in terms of $f(\phi_-)$~, such that the variational principle is well defined with all terms being intrinsic to the boundary. We leave this as an interesting open problem for the future. 

These results indicate that hairy accelerating black holes can be used to study holographic CFTs with marginal multi-trace deformations with a non-constant holographic scalar operator. Similar structures have previously been found \cite{deHaro:2006wy, deHaro:2006ymc, Papadimitriou:2007sj} but they correspond to instanton solutions, namely, they have no backreaction on the metric tensor and a trivial stress tensor, i.e., stealth configurations. They serve to study the decay of the conformal vacuum of the dual theory as these configurations correspond to extrema of the scalar potential of the dual theory.

\section{Other geometries: Asymptotically locally de Sitter and asymptotically locally flat solutions}

Through this paper, we have focused on the physics of AlAdS solutions. Notwithstanding, a remarkable feature of our class of solutions \eqref{SOL} is that it provides us with asymptotically locally dS and asymptotically locally flat geometries as well. Moreover, these exist in both sectors, $\sigma=-1$ and $\sigma=1$~. While we do not intend to provide an exhaustive analysis of these spacetimes here,   we shall construct illustrative cases in order to understand the novelty and main properties of these spacetimes.

\begin{figure}[H]
    \centering
    \includegraphics[scale=1]{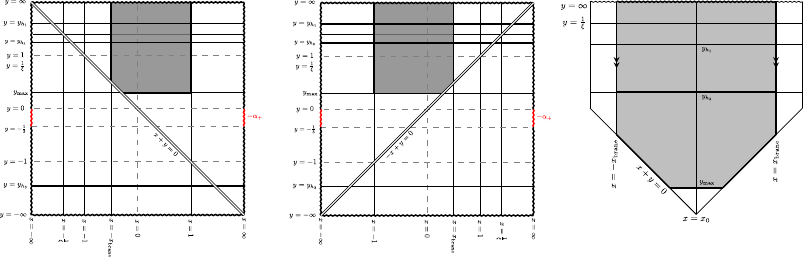}
    \caption{Asymptotically locally dS slowly accelerating black hole. This geometry represents the spacetime of a slowly accelerating black hole with event and cosmological horizons. It accelerates due to the presence of one domain wall and it belongs to the family $\sigma=1$~. Here $0<\xi\leq1$~.}
    \label{fig:side_by_side11}
\end{figure}

An appealing case is the dS geometry in the $\sigma=1$ family when $0<\xi<1$~. As shown in the appendix, the generic case contains three Killing horizons satisfying $y_{h_3}<-1<0<1<y_{h_2}<y_{h_1}<\frac{1}{\xi}$~. An extremal case arises when $y_{h_2}$ and $y_{h_1}$ coincide with the local maximum of the function $f_1(y)$~. Observe that the scalar field profile will always be real and well-behaved whenever the $\alpha_+$ branch is considered and the $y$-coordinate satisfies $y_{\rm max}>0\geq-\alpha_+$~.
This spacetime, depicted in \autoref{fig:side_by_side11}, exhibits an accelerating dS black hole supported by a single domain wall located at the left hand side of $x_{\rm max}=1$~. 

\begin{figure}[h]
    \centering
    \includegraphics[scale=1]{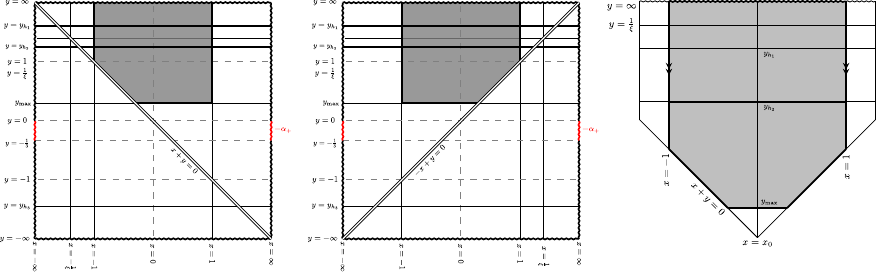}
    \caption{ Asymptotically locally dS black hole. This geometry represents the spacetime of a locally dS black hole with event and cosmological horizons. No topological defect is included, reason why it is believed to not represent an accelerating black hole. Belongs to the family $\sigma=1$ with $0<\xi\leq1$~.}
    \label{fig:side_by_side12}
\end{figure}

It is slowly accelerating and it features a single event horizon dressed by a cosmological horizon. However, notice that in this case the identification of the transverse coordinate can also be performed without introducing the brane. In fact, both points $x_{\rm min}$ and $x_{\rm max}$ are tensionless and in consequence, no effective tension-full brane would take place. This solution represents a non-accelerating hairy three-dimensional dS black hole, and it is novel in its own right. See \autoref{fig:side_by_side12}.

We close this section with some comments on the asymptotically locally flat case. In this case, the distribution of the Killing horizon is accessible analytically. It simply follows from the study of the Killing horizons of the asymptotically locally flat four-dimensional C-metric. However, in contrast with the four-dimensional case, in three dimensions the interval of the transverse coordinate substantially modifies the final form of the spacetime. 
We again focus on the $\sigma=1$ family, as it produces an interval $(x_{\rm min},x_{\rm max})$ of which both extremes are tensionless points. Regardless of the choices $0<\xi\leq1$ or $\xi>1$~, we always find a geometry with two Killing horizons. 

\begin{figure}[h]
    \centering
    \includegraphics[scale=1]{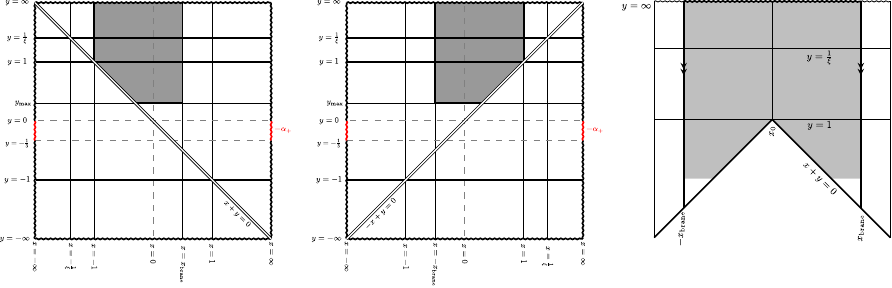}
    \caption{Asymptotically locally flat rapidly accelerating black hole. This geometry represents the spacetime of a rapidly accelerating black hole with event and Rindler horizons. It accelerates due to the presence of one domain wall and it belongs to the family $\sigma=1$~. Here $0<\xi\leq1$~.}
    \label{fig:side_by_side13}
\end{figure}

\begin{figure}[h]
    \centering
    \includegraphics[scale=1]{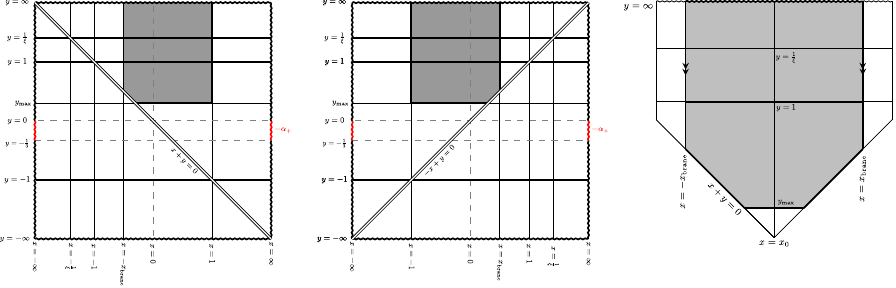}
    \caption{Asymptotically locally flat slowly accelerating black hole. This geometry represents the spacetime of a slowly accelerating black hole with inner and event horizons. It accelerates due to the presence of one domain wall and it belongs to the family $\sigma=1$~. Here $0<\xi\leq1$~.}
    \label{fig:side_by_side14}
\end{figure}

The nature of the horizons depends on the localization of the brane. Either one event and one Rindler horizon are present or an outer/inner event horizon pair appears.
 Thus, rapidly accelerating or slowly accelerating asymptotically locally flat black holes accelerated by one domain wall can be always achieved. The scalar field profile will be always well-behaved as long as we 
chose the $\alpha_+$ branch and impose $y_{\rm max}>0\geq-\alpha_+$; we illustrate the possibilities in \autoref{fig:side_by_side13} and \autoref{fig:side_by_side14}.

\begin{figure}[H]
    \centering
    \includegraphics[scale=1]{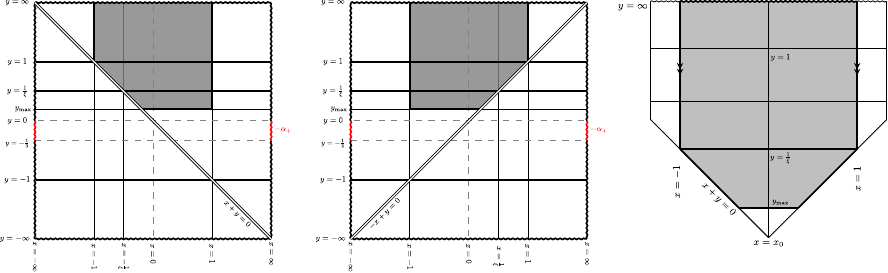}
    \caption{Asymptotically locally flat black hole. This geometry represents the
spacetime of an asymptotically locally flat black hole with one event horizon dressed by a non-compact horizon. No brane is present whatsoever reason why it is believed to not possess any acceleration. It exists in the family $\sigma=1$~. Here $\xi>1$~.}
    \label{fig:side_by_side15}
\end{figure}

We also note that, as in the previous dS case, a solution with no brane whatsoever also exists. This is depicted in \autoref{fig:side_by_side15}, which represents an asymptotically locally flat solution with one event horizon and a non-compact horizon. This case merits a more detailed investigation,  as here the brane is not included, and therefore there is no mechanism for the black hole to accelerate.

\newpage

\section{Conclusions}

Our investigation of the construction of 
hairy accelerating three-dimensional black holes has resulted in a broad set of solutions falling into several classes. 
These were found by considering the backreaction of a conformally coupled self-interacting scalar field. The solutions we have obtained are novel, having never before appeared in the literature. 

Our starting point was a pedagogical review of the foundational principles underlying the construction of vacuum three-dimensional accelerating solutions  \cite{Astorino:2011mw, Arenas-Henriquez:2022www}. We then extended this methodology to incorporate a conformally coupled scalar field. In this scenario, the tension of the domain wall is computed via a generalized version of the Israel junction conditions, namely those that also consider the jump of the scalar field along the wall. The causal structure of any given solution is therefore crucially determined by the location of the domain wall, the restriction imposed by an everywhere real and well-behaved scalar field profile, and the value of the corresponding cosmological constant. 

Our primary focus was on studying black holes. These geometries are found to possess up to three Killing horizons:  inner, outer (event),  and accelerating horizons, and their specific distribution gave rise to interesting rapidly and slowly accelerating spacetimes. We paid particular attention to AlAdS geometries, not only for their intrinsic relevance in three-dimensional gravity but also for their applications in holography. We have obtained an AlAdS accelerating black hole with event and Rindler horizons and an AlAdS slowly accelerating black hole with inner and event horizons,  generalizing the accelerating extension of the BTZ black hole that was recently obtained \cite{Astorino:2011mw, Arenas-Henriquez:2022www}. The whole hierarchy of spacetimes displayed from our set of solutions can be seen in \autoref{fig:hierarchiessm1}. Along the same lines, we constructed an AlAdS rapidly accelerating black hole with inner, event, and Rindler horizons and an AlAdS slowly accelerating black hole with three compact horizons, that generalize the accelerating particle geometries and the accelerating vacuum black hole with no BTZ limit found previously in \cite{Arenas-Henriquez:2022www}. This hierarchy of solutions is found in \autoref{fig:hierarchiessp1}.

We then conducted holographic studies of these hairy AlAdS black holes. 
First, we considered a new set of coordinates that allowed us to properly understand the asymptotic behavior of the fields. The theory on-shell is equivalent to AdS gravity with a minimally coupled scalar exhibiting a hexic self-interaction and a specific mass term, the latter of which is consistent with the BF bound. Following this, we built the renormalized on-shell action, along with its associated holographic stress tensor that was used to compute the black hole mass. We found that the domain wall modifies the action principle and serves as a counterterm to address additional divergences that appear due to the acceleration, just as in the vacuum case \cite{Arenas-Henriquez:2023hur}. 

The holographic stress tensor revealed several features worth mentioning. First, its trace recovers the two-dimensional Weyl anomaly. Second, it is not covariantly conserved such that one can identify a holographic Ward identity. Finally, it can be cast in the form of the stress tensor of a perfect fluid within a curved background. We have also elaborated on how the boundary CFT contains marginal multi-trace deformations when the scalar field satisfies mixed boundary conditions. 

Last, but not least, we constructed geometries with asymptotically locally dS and asymptotically locally flat behavior. These solutions are interesting in their own right, as they are novel in the three-dimensional setting. In the dS case, we found a slowly accelerating black hole with event and cosmological horizons. We also found a dS geometry with event and cosmological horizons and with no domain wall whatsoever. This spacetime is therefore preliminarily categorized as non-accelerating. Along the same lines, we constructed an asymptotically locally flat rapidly accelerating black hole with event and Rindler horizons and an asymptotically locally flat slowly accelerating black hole with inner and event horizons. As in the dS case, a geometry with no brane whatsoever was also identified;  it can be regarded as an asymptotically locally flat black hole with one event and one non-compact horizon. The properties of this object warrant further study, due to the fact that in the absence of a domain wall, the solution is not expected to possess acceleration. 

Several future research directions emerge from our work. First, it would be desirable to  investigate  the thermodynamic properties of these accelerating black holes,
explicating in particular
their distinctions from the BTZ case 
\cite{Frassino:2015oca},
 the Smarr relation, the isoperimetric inequality \cite{Cvetic:2010jb},  or the thermodynamic decay of the hairy accelerating black hole to the bald case, in analogy to the thermal decay of the HMTZ black hole into the BTZ black hole \cite{Park:2004yk, Gegenberg:2003jr}. 
Recently it has been argued that the action \eqref{action1} can be rewritten, modulo boundary terms, as the difference of two Chern--Simons actions with composite gauge connections evaluated over a Lie algebra that depends on the value of the cosmological constant and the sign of the self-interacting potential \cite{Cardenas:2022jtz}. 
The thermodynamic properties of the HMTZ black hole can then be studied by using the conserved charges and holonomy conditions along the gauge fields. It would be interesting to use such a framework to obtain the thermodynamic charges of the hairy accelerating black holes we have obtained.

Furthermore, several other aspects of accelerating black hole thermodynamics are yet not fully understood,  especially in three dimensions but not exclusively. The first law of four-dimensional accelerating black holes can be extended by considering variations of the cosmic string tensions \cite{Anabalon:2018ydc}. Nonetheless, varying the tension may induce a change in the topology of the solution, and it has been shown that the first law can be obtained with no need for such terms \cite{Appels:2016uha, Cassani:2021dwa}. In addition, it has been argued that, in order for the first law to be satisfied, one needs to introduce a rescaling of the time coordinate \cite{Anabalon:2018ydc}, which is fixed such that the massless black hole can be mapped to the Rindler-AdS solution, which corresponds to the accelerating vacuum. However, it was recently shown that more than one rescaling leads to having a first law \cite{Kim:2023ncn}. 

Finally, three-dimensional AdS gravity is afflicted by a holographic Weyl anomaly. Consequently, the renormalized Euclidean action possesses on-shell conformal symmetry except
for this anomalous term. This leads to a subtle distinction in the first law of thermodynamics for AlAdS black holes in odd
dimensions \cite{Papadimitriou:2004ap}. When considering arbitrary variations of the thermodynamic quantities, the
black hole energy does not remain invariant under Weyl transformations. To address this, a compensating Penrose-Brown-Henneaux transformation \cite{Penrose:1985bww, Brown:1986nw} must be introduced in the bulk, which is responsible for Weyl rescaling at the boundary. This translates to the fact that for odd-dimensional AlAdS, the first law depends on the conformal representative of the boundary metric, introducing a variation in the Casimir energy of AdS.  As explained in Section 3,
the upper-half plane coordinates $(z,\tau,x)$ given in \eqref{Hologzcanonical}, suffice to compute holographic quantities using the holographic stress tensor. However, as clarified in \cite{Hubeny:2007xt}, these coordinates only match the Fefferman–Graham gauge at the leading order. Thus, it remains unclear whether a compensating transformation is necessary to establish homogeneity with the first law of black hole mechanics, see \cite{Arenas-Henriquez:2023vqp} for a detailed analysis.

All the subtleties mentioned above render the task of deriving a first law for accelerating black holes in three dimensions quite challenging. We believe that reproducing the quantum statistical relation for these black holes would be a significant step in establishing
that this system is indeed in thermal equilibrium. Nevertheless, the full understanding of the
first law remains an intriguing open problem for the future.

A crucial step in understanding the holographic description of three-dimensional accelerating black holes is to embed the solutions into string theory. In \cite{Ferrero:2020twa} it was shown that by restricting the physical parameter of the AdS Pleba\'nski--Demia\'nski (PD) family of solutions\footnote{The NUT parameter is not considered in this discussion.}, it can be uplifted to a seven-dimensional Sasaki--Einstein space to obtain a local solution to $D=11$ supergravity. 
The construction of the higher-dimensional solution relies on the fact that the four-dimensional solution is a solution of gauged supergravity, whose bosonic sector consists simply of AdS Einstein--Maxwell theory, and on the choice of a particular parameter space where the conical singularities of the PD black hole are quantized. The horizon then becomes a spindle. As any solution of four-dimensional gauged supergravity can be uplifted to an arbitrary seven-dimensional Sasaki--Einstein manifold, to find a solution of eleven-dimensional supergravity \cite{Gauntlett:2007ma}, the authors of \cite{Ferrero:2020twa} showed that the PD solution with restricted parameters can be embedded into supergravity in eleven dimensions by means of a regular seven-dimensional Sasaki--Einstein manifold.
The elevated solution is regular, namely, it is free of conical singularities, with quantized fluxes that are related to the spindle data of the lower-dimensional solution. 
The process of uplifting three-dimensional accelerating solutions to supergravity remains an open problem; it is a crucial step toward understanding the underlying description of CFTs dual to AlAdS solutions with acceleration. For recent advancements in research regarding the connection between the C-metric and conformal symmetry, see  \cite{Lei:2023mqx, Xue:2023lid}.

Another particularly interesting direction is to have better control of the boundary geometry. As we have considered new coordinates that match the Fefferman--Graham gauge only at first order, which is enough to have some relevant holographic quantities, a full understanding of the holographic theory has yet to be achieved. 
As the coordinates \eqref{Hologzpolar}, \eqref{Hologzcanonical} induce cross terms between the holographic coordinate $z$ and boundary coordinates $x^i$~, a good candidate for carrying out a full analysis of the holographic theory is to consider the 
Fefferman--Graham--Weyl gauge \cite{Ciambelli:2019bzz, Jia:2021hgy, Jia:2023gmk}. This gauge is used to restore Weyl covariance at the boundary and considers cross terms of the holographic and boundary coordinates. Moreover, the diffeomorphism mapping the Weyl--Fefferman--Graham gauge to the Fefferman--Graham gauge can be charged, and thus non-trivial \cite{Ciambelli:2023ott}.  This implies that the Fefferman--Graham gauge, although is always achievable, may restrict the moduli space of the bulk gravitational theory. 
 Holographic data has recently been constructed in this gauge
\cite{Ciambelli:2023ott},
including the covariant Weyl connection, the holographic renormalization procedure, and new conserved charges associated with residual gauge symmetries. Particularly, the first order in the asymptotic expansion of the metric in the Weyl--Fefferman--Graham gauge is mapped to a Weyl connection in the boundary, and sources a new boundary current. 
We plan to move forward on the holographic analysis of these three-dimensional accelerating solutions by considering this framework.

\acknowledgments

We are grateful to G. Arenas--Henriquez and David Rivera--Betancour for helpful discussions and to Jos\'e Barrientos for participation on an early stage of this work.
The work of A.C. is funded by Primus grant PRIMUS/23/SCI/005 from Charles University and FONDECYT Regular grant No. 1210500.
The work of F.D. is supported by {\sc Beca Doctorado nacional} 
(ANID) 2021 Scholarship No. 21211335, ANID/ACT210100 Anillo Grant ``{\sc Holography and its applications to High Energy Physics, Quantum Gravity and Condense Matter Systems}'' and FONDECYT Regular grant No. 1210500. J.O. is partially supported by FONDECYT Grant 1221504. 
R.B.M. is supported in part by the Natural Sciences and Engineering Research Council of Canada. A.C. would like to express his gratitude to the International Centre For Theoretical Physics (ICTP), for providing such a wonderful environment under which this work has been finished.  

\begin{appendix}

\section{Killing horizons}\label{App:KillingHorizons}

A crucial first step in analyzing the geometries is to identify the black hole horizons. Following this, a meticulous definition of the spacetime coordinates becomes possible.
The positioning of the Killing horizons is determined by the condition
\begin{equation}
    F(y)=-\sigma(1-y^2)(1-\xi y)+\frac{1}{A^2\ell^2}=0~.
\end{equation}
Due to the cubic nature of the polynomials and the influence of the cosmological constant, the localization of the Killing horizons becomes an algebraically involved process. Nonetheless,  by employing simple numerical techniques, we can extract essential information that sufficiently reveals the horizon structure of the solutions.

To begin, we define the two auxiliary functions
 \begin{equation}
f_0=\frac{1}{A^2 \ell^2}, \hspace{0.5cm}  f_1=-\sigma(1-y^2)(1-\xi y)~, 
\end{equation}
whose intersection provides qualitative information regarding the position of the Killing horizons of $F(y)$~. First, the function $f_0$ is constant. Its absolute value depends on the value of the acceleration (assuming the value of $l$ is fixed). For $A<<1$~, $|f_0|$ will always acquire large values while for very large $A$ it will always approach zero. The sign of $f_0$ depends on the sign of the cosmological constant. In AdS, $\ell\in\mathbb{R}$~, and thus $f_0$ is always negative. On the other hand, for dS, the situation is the opposite, $\ell\in\mathbb{C}$~, and so $f_0$ is positive. 

The function $f_1$ on its own has a richer structure. It  is not constant and   features three real roots,  $-1$~, $1$ and $1/\xi$~, as well as a local maximum and local minimum located at 
\begin{equation}
    y_{{\rm{max}}/{\rm min}}=\frac{1\pm\sqrt{1+3\xi^2}}{3\xi}~.
\end{equation}
These features are independent of the value of $\sigma$ and therefore are valid for all the families of solutions we consider in this paper. In spite of this, the asymptotic behavior of $f_1$   depends on the value of $\sigma$~. Hence, the global analysis of $f_1$ requires independent analysis for each of the cases $\sigma=\pm1$~. 

We begin with $\sigma=-1$~. In this case large positive values of $y$ make the function $f_1$ to tend to plus infinity. In a similar fashion, large negative values of $y$ tend  $f_1$ toward minus infinity. At the origin the function satisfies $f_1(0)=1$~. The case $\sigma=1$ exhibits the opposite behavior, namely for large values of $y$ the function $f_1$ approaches minus infinity, whereas it approaches plus infinity for large negative values of $y$~. In addition, $f_1(0)=-1$~. 

\begin{figure}[htbp]
  \begin{minipage}[t]{0.45\linewidth}
    \centering
\includegraphics[scale=0.37]{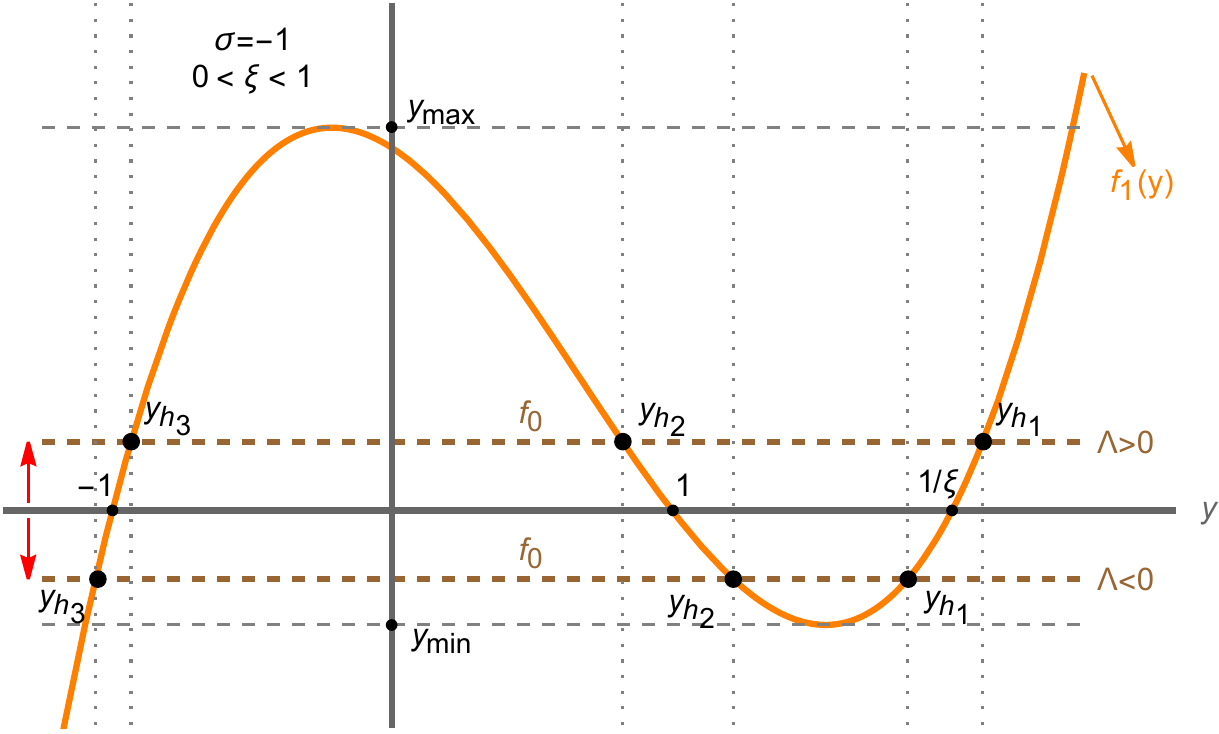}
  \end{minipage}
  \hfill
  \begin{minipage}[t]{0.45\linewidth}
    \centering
   \includegraphics[scale =0.37]{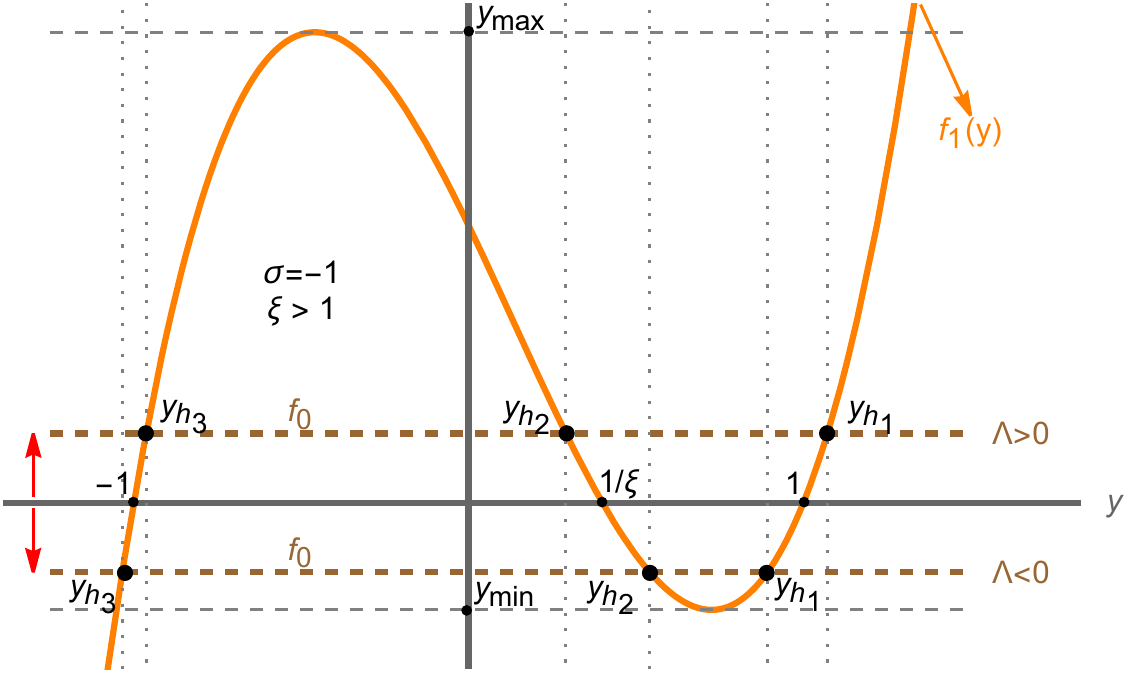}
  \end{minipage}
  \caption{$\sigma=-1$}
  \label{horsminus1}
\end{figure}

\begin{figure}[htbp]
  \begin{minipage}[t]{0.45\linewidth}
    \centering
\includegraphics[scale=0.37]{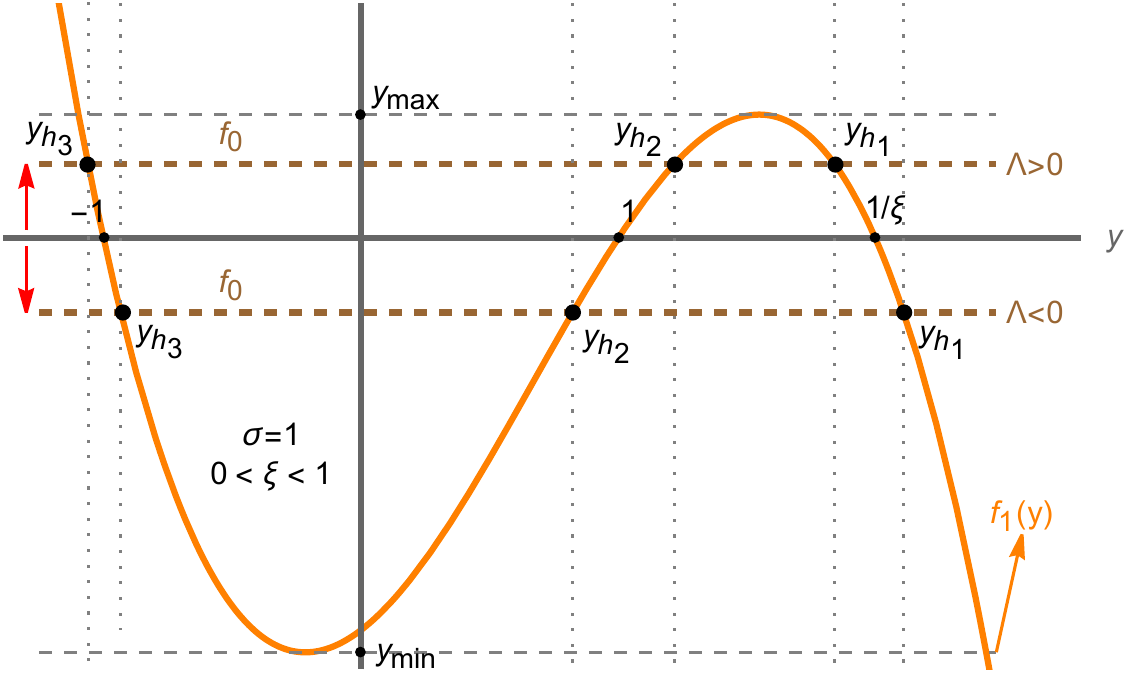}
  \end{minipage}
  \hfill
  \begin{minipage}[t]{0.45\linewidth}
    \centering
   \includegraphics[scale =0.37]{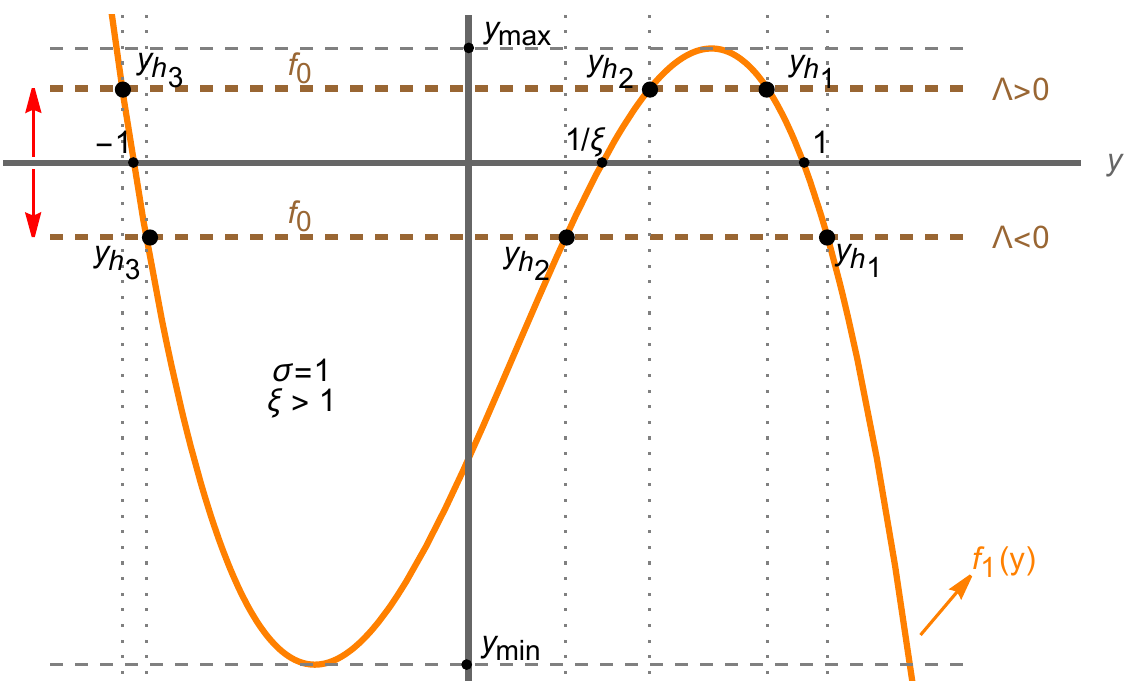}
  \end{minipage}
  \caption{$\sigma=1$}
  \label{horplus1}
\end{figure}

With this information at hand, we can effectively study the intersection of the functions $f_0$ and $f_1$ and therefore qualitatively achieve the position of the Killing horizons. This is graphically expressed in \autoref{horsminus1} and \autoref{horplus1}.

The qualitative analysis we have performed allows us to identify the distribution of the Killing horizons in all cases of interest: for the families $\sigma=\pm1$ with both signs of $\Lambda$ and for a positive value of the parameter $\xi$~, $0<\xi<1$ or $\xi>1$~. The following scenarios are extracted.
\subsection{$\sigma=-1$ and $\Lambda<0$}
The most general scenario contains three Killing horizons. Starting with $0<\xi<1$ we observe that 
\begin{equation}
   y_{h_3}<-1<0<1<y_{h_2}<y_{h_1}<\frac{1}{\xi}~.
\end{equation} 
Two different configurations can be achieved for small enough values of the acceleration parameter $A$~. First, if $f_0$ coincides with the local minimum $y_{\rm min}$~,  the horizons $y_{h_2}$ and $y_{h_1}$ merge into a single degenerate horizon within the interval $(1,\frac{1}{\xi})$~. Second, if $f_0<y_{\rm min}$ then only $y_{h_3}$ prevails and the geometry features a single Killing horizon. This horizon structure is, in practical terms, basically maintained when $\xi>1$~. Here we have 
\begin{equation}
   y_{h_3}<-1<0<\frac{1}{\xi}<y_{h_2}<y_{h_1}<1~.
\end{equation} 
Both exceptional cases maintain their features. 

\subsection{$\sigma=-1$ and $\Lambda>0$}

For a positive cosmological constant, we encounter a positive function $f_0$~. The general case with three Killing horizons now implies
\begin{equation}
   -1<y_{h_3}<y_{h_2}<0<1<\frac{1}{\xi}<y_{h_1}~,
\end{equation} 
when $0<\xi<1$~. In this case, if $f_0$ coincides with the local maximum $y_{\rm max}$ we have an extremal case in which $y_{h_3}$ and $y_{h_2}$ merge into a single degenerate horizon lying in the interval $(-1,0)$~. Finally, for sufficiently small values of $A$ only the Killing horizon $y_{h_1}$ prevails in the geometry. An analogous situation holds when $\xi>1$~. The Killing horizons now respect 
\begin{equation}
   -1<y_{h_3}<y_{h_2}<0<\frac{1}{\xi}<1<y_{h_1}~.
\end{equation} 
The same exceptional cases are found for small values of the acceleration. 

\subsection{$\sigma=1$ and $\Lambda<0$}

For $0<\xi<1$ we observe the general case 
\begin{equation}
   -1<y_{h_3}<0<y_{h_2}<1<\frac{1}{\xi}<y_{h_1}~.
\end{equation} 
Here, an extremal case arises if $f_0$ coincides with the local minimum $y_{\rm min}$~, where $y_{h_3}$ and $y_{h_2}$ merge in a degenerated horizon within the interval $(-1,0)$~. A single horizon geometry arises for even smaller values of $A$ in which only $y_{h_1}>\frac{1}{\xi}$ remains. Similarly for $\xi>1$ we get 
\begin{equation}
   -1<y_{h_3}<0<y_{h_2}<\frac{1}{\xi}<1<y_{h_1}~.
\end{equation} 
Both exceptional cases, the extremal case, and the single horizon case follow the same structure, now with $y_{h_1}>1$~. 

Notice that in both cases $y_{h_2}$ is positive unless the numerical value of the cosmological constant is large enough for $y_{h_2}$ to cross $y=0$~. This is the previous stage to the merging of $y_{h_2}$ and $y_{h_3}$ into the extremal case

\subsection{$\sigma=1$ and $\Lambda>0$}

In this family, a positive cosmological constant and a $\xi$ parameter satisfying $0<\xi<1$ yield the general case 
\begin{equation}
y_{h_3}<-1<0<1<y_{h_2}<y_{h_1}<\frac{1}{\xi}~.
\end{equation} 
An extremal case arises when $f_0$ reaches the local maximum $y_{\rm max}$~. $y_{h_2}$ and $y_{h_1}$ merge into a single degenerate horizon within the interval $(1,\frac{1}{\xi})$~. Moving to smaller values of $A$ provides the case with one single Killing horizon $y_{h_3}<-1$~. Similarly, for $\xi>1$ we have 
\begin{equation}
y_{h_3}<-1<0<\frac{1}{\xi}<y_{h_2}<y_{h_1}<1~.
\end{equation} 
The same extremal and single horizon cases emerge for a properly restricted small enough value of the acceleration. \\

On the basis of these geometries, several spacetimes can be constructed by wisely choosing the set $(x_{\rm min},x_{\rm max})$ in which the transverse coordinate lies. 
\end{appendix}

\newpage
\bibliographystyle{JHEP}
\bibliography{biblio}
\end{document}